\documentclass{jfm}
\usepackage{graphicx}
\usepackage{epstopdf, epsfig}
\usepackage{bm}
\usepackage{amsmath}
\usepackage{enumerate}
\usepackage{xcolor}

\shorttitle{Collision-Induced breakage of agglomerates in turbulence}
\shortauthor{S. Chen et al.}

\title{Collision-induced breakage of agglomerates in homogenous isotropic turbulence laden with adhesive particles}

\author{Sheng Chen\aff{1}
  \corresp{\email{sheng\_chen@hust.edu.cn}},
  Shuiqing Li\aff{2}
}

\affiliation{
\aff{1}State Key Laboratory of Coal Combustion, School of Energy and Power Engineering, Huazhong University of Science and Technology, Wuhan 430074, China

\aff{2}Key Laboratory for Thermal Science and Power Engineering of Ministry of Education, Department of Energy and Power Engineering, Tsinghua University, Beijing 100084, China
}

\begin{document}

\maketitle

\begin{abstract}
We carry out direct numerical simulation combined with adhesive discrete element calculations (DNS-DEM) to investigate collision-induced breakage of agglomerates in homogeneous isotropic turbulence. The adopted method tracks the dynamics of individual particles while they are travelling alone through the ﬂuid and while they are colliding with other particles. Based on extensive simulation runs, an adhesion parameter $\rm{Ad}_n$ is constructed to quantify the possibility of occurrence of sticking, rebound, and breakage events. The collision-induced breakage rate is then formulated based on the Smoluchowski equation and a breakage fraction. The breakage fraction, defined as the fraction of collisions that result in breakage, is then analytically estimated by a convolution of the probability distribution of collision velocity and a universal transfer function. It is shown that the breakage rate decreases exponentially as the adhesion parameter $\rm{Ad}_n$ increases for doublets and scales as linear functions of the agglomerate size, with the slope controlled by $\rm{Ad}_n$. These results allow one to estimate the breakage rate for early-stage agglomerates of arbitrary size. Moreover, the role of the flow structure on the collision-induced breakage is also examined. Violent collisions and breakages are more likely caused by particles ejected rapidly from strong vortices and happen in straining sheets. Our results extend the findings of shear-induced fragmentation, forming a more complete picture of breakage of agglomerates in turbulent flows.
\end{abstract}


\section{Introduction}
For solid micron particles immersed in turbulence, various complicated particle-scale interactions, such as van der Waals attraction \citep{Israelachvili2011, ChenPRF2019}, capillary force \citep{RoyerNature2009}, and electrostatic forces \citep{Jones2005,SteinpilzNP2019}, lead to the formation of agglomerates. On the other hand, breakage of agglomerates also happens due to the flow stress \citep{HigashitaniCES2001, bablerJFM2008} and collisions of other particles \citep{LiuPRL2018}. Both the formation and the breakage of agglomerates find broad applications in industry, ranging from particulate matter control \citep{ChangFuel2017, JaworekPECS2018, WeiNED2019}, drug delivery \citep{VossIJP2002}, agglomerate dispersion in gas phase \citep{IimuraAPT2009} to water treatment \citep{RenaultCEJ2009}. However, to predict if and how fast agglomeration and deagglomeration occur in turbulence is highly challenging because of the multi-scale characteristics associated with both turbulent flows and the interacting modes between particles \citep{MarshallJCP2009, LiPECS2011, Marshall2014}.

The mechanisms of agglomeration have been extensively studied. It is generally accepted that the turbulent ﬂow first brings two initially separate particles at a sufﬁciently close distance, and microphysical mechanisms (collisional dissipation, hydrodynamic interactions, surface effects) then determine whether the two approaching particles can form an agglomerate. Collision kernels, expressed as the product of the mean relative radial velocity and the radial distribution function, have been proposed to predict the rate at which the flow brings separate particles into contact \citep{SaffmanJFM1956, WangJFM2000}. The kernel functions are further extended to reﬂect the inﬂuence of particle inertia, identifying the effect of preferential concentration \citep{SquiresPOF1991, SawPRL2008, BalachandarAR2010, TagawaJFM2012} leading to an inhomogeneous particle distribution and sling or caustic effects \citep{FalkovichNature2002, WilkinsonPRL2006, pumir2016}, which cause inertial particles to collide with large velocity differences. Recent studies also suggest that complicated interparticle interactions, including elastic repulsion \citep{BecPRE2013, VosskuhlePRE2013}, electrostatic interactions \citep{lu2010clustering, lu2015charged} and van der Waals adhesion \citep{ChenPRF2019, KelloggJFM2017}, give rise to nontrivial collision phenomenon that cannot be predicted from the \textit{ghost collision approximation}, where particles can pass through each other without any modiﬁcation to their trajectories.

The breakage of agglomerates, in contrast, is still far from clear. Previous studies mainly focus on \textit{shear-induced breakage}. Discrete particle approach, which provides information at the particle level, has been employed to better understand the relationship between flow strain rate and the internal stress of agglomerates. For isostatic agglomerates exposed to the flow, the forces and torques on each elementary particle can be calculated assuming force and torque balances on all particles \citep{SetoPRE2011, VanniLangmuir2011, FellayJCIS2012}. The bond between particles instantly breaks up if the interparticle force reaches a critical value (bond strength), leading to the breakage of the isostatic agglomerate \citep{BonaJFM2014, BablerJFM2015}. To simulate the breakage of hyperstatic agglomerates with a dense structure, soft-sphere discrete element method (DEM) is usually regarded as a powerful tool. In DEM, translational and rotational motions of all particles in an agglomerate are integrated with a sufﬁciently small time step so that the deformations at the contact region are resolved. Based on DEM simulations, a criterion for shear-induced breakage of hyperstatic agglomerates has been proposed, which is valid across a wide range of shear stress and interparticle adhesion values \citep{RuanCES2020}.


\textcolor{black}{
Turbulent flows are usually considered to enhance the clustering and agglomeration of particles. However, recent work has revealed that a stronger clustering effect gives rise to a higher collision velocity, which increases the breakage rate of agglomerates \citep{LiuPRL2018}. The collision-induced breakage is important for gas-solid systems containing small but heavy particles (with high Stokes numbers). Such systems exist in the electrostatic agglomerators for the removal of ﬂy ash particles from ﬂue gas \citep{JaworekPECS2018}, gas-cooled reactors containing graphite aerosols \citep{WeiNED2019}, and ﬂuidized beds with Geldart Group A particles \citep{gu2016modified}. The competition between clustering and deagglomeration provides an explanation for the saturation of agglomeration levels in these gas-solid systems.
} To predict the kernel function for collision-induced breakage in turbulence requires one to know (i) the statistics of particle collision velocity; (ii) the particle-scale interactions (e.g., adhesions, elastic repulsions, and frictions), which determine whether two colliding agglomerates will either merge into a large one, rebound from each other or break up into fragments \citep{Dizaji2019JFM}. However, to our knowledge, the formulation of the breakage rate that can reflect both these two aspects is still far from perfect. Besides, it has been suggested that flow structure significantly affect the collisions of \textit{non-interacting particles} \citep{bec2016abrupt, PicardoPRF2019, XiongIJMF2019}. It is not clear how to correlate the collision-induced breakage to the structure of flows.

In this paper, we try to address the issues above by investigating the collision-induced deagglomeration of solid adhesive particles in homogeneous isotropic turbulence (HIT). An adhesive DEM is employed to fully resolve the translational and rotational motions of all particles. We first introduce how to identify various events, including sticking, rebound, collision-induced breakage and shear-induced breakage of agglomerates, in simulations. The collision-induced breakage rate is then formulated based on the Smoluchowski equations and a breakage fraction function. A universal transfer function is proposed to predict the breakage fraction function from the probability distribution of collision velocity.
We also demonstrate how intense vorticity and strain contribute to the breakage of agglomerates and show how the breakage rate scales with particle size, particle number density and agglomerate size.

\section{Methods}
\subsection{Fluid phase calculation}
To investigate the collision-induced breakage of agglomerates, we consider non-Brownian solid particles suspended in an incompressible isotropic turbulent flow, which is calculated by DNS on a cubic, triply-periodic domain with $128^3$ grid points. A pseudospectral method with second-order Adams-Bashforth time stepping is applied to solve the continuity and momentum equations
\begin{subequations}
\begin{align}
  \nabla \cdot {\bm u} &= 0, \label{eq:fcont} \\
  \frac{\partial \boldsymbol{u}}{\partial t}+(\boldsymbol{u} \cdot \boldsymbol{\nabla}) \boldsymbol{u} &=-\boldsymbol{\nabla} p+\nu \nabla^{2} \boldsymbol{u}+{\bm f}_F +{\bm f}_P. \label{eq:fmom}
\end{align}
\end{subequations}
Here, ${\bm u}$ is the fluid velocity, $p$ is the pressure, $\rho_f$ is the fluid density, and $\nu$ is the fluid kinematic viscosity. The small wavenumber forcing term ${\bm f}_F$ is used to maintain the turbulence with an approximately constant kinetic energy. ${\bm f}_P$ is the particle body force, which is calculated at each Cartesian grid node $i$ using ${\bm f}_p({\bm x}_i) \!=\! - \sum_{n \!=\! 1}^N {\bm F}_n^F \delta_h \left({\bm x}_i \!-\! {\bm X}_{p,n} \right)$. Here, ${\bm x}_i$ is the location of grid node $i$, ${\bm F}_n^F$ is the fluid force on particle $n$ located at ${\bm X}_{p,n}$ and $\delta_h \left({\bm x}_i \!-\! {\bm X}_{p,n} \right)$ is a regularized delta function, which is given by
\begin{equation}
\label{delta_twoway}
\delta_{h}\left({\bm x}_{i}-X_{p, n}\right)=\left\{\begin{array}{cc}\frac{n_{b i}}{N_{g} N_{b}} & \text { if } {\bm x}_{i} \in \mathcal{N}^{B} \\ 0 & \text { if } {\bm x}_{i} \notin \mathcal{N}^{B}\end{array}\right.
\end{equation}
\textcolor{black}{Here, $\mathcal{N}^B$ is the set consisting of the grid cell containing the particle and one grid cell on each side, $N_b=27$ is the number of grid cells that in the set $\mathcal{N}^B$, $N_g=8$ is the number of grid nodes in each grid cell, and $n_{bi}$ is the number of grid cells in the set $\mathcal{N}^B$ that containing the grid node $x_i$. The summation of $\delta_h$ over all grid nodes is unity, i.e., $\sum_{{\bm x}_{i}} \delta_{\mathrm{h}}\left({\bm x}_{i}-X_{p, n}\right)=1$, indicating that the choice of delta function is conservative in force.}

\textcolor{black}{All the parameters in our simulation have been nondimensionalized by typical length, velocity, and mass scales that are relevant to the agglomeration of microparticles. Specifically, the typical length scale is $L_0 = 100r_p = 0.001 \rm{m}$, where $r_p = 10 \rm{\mu m}$ is the particle radius. The velocity scale is set as $U_0 = 10 \rm{m/s}$ which is the typical value for the gas flow in a turbulent-mixing agglomerate \citep{JaworekPECS2018}. The typical mass is $M_0 = \rho_f L_0^3= 10^{-9} \rm{kg}$, where $\rho_f=1 \rm{kg/m^3}$ is the ﬂuid density. The typical timescale is given by $T_0=L_0/U_0=10^{-4} \rm{s}$. Other dimensional input parameters are the ﬂuid viscosity $\mu =1.0\times 10^{-5} \rm{Pa\cdot s}$, the particle density $\rho_p = 10 \sim 320 \rm{kg/m^3}$, and the particle surface energy $\gamma = 0.01\sim5 \rm{J/m^2}$. Hereinafter, all the variables appear in their dimensionless form and, for simplicity, the same notations as the dimensional variables are used. One could obtain “physical” values of dimensionless variables by multiplying the dimensionless values with the typical scales.}

\subsection{Equations of motion and particle-particle interactions}
A soft-sphere DEM is employed to track the dynamics of every individual particle. We integrate the linear and angular momentum equations of particles
\begin{subequations}
\begin{align}
  m_i \dot{\bm v}_i = {\bm F}_i^F + {\bm F}_i^C, \label{eq:eoma} \\
  I_i \dot{\bm \Omega}_i = {\bm M}_i^F + {\bm M}_i^C. \label{eq:eomb}
\end{align}
\end{subequations}
where $m_i$ and $I_i$ are mass and moment of inertia of particle $i$ and ${\bm v}_i$ and ${\bm \Omega}_i$ are the translational velocity and the rotation rate of the particle. The forces and torques are induced by both the fluid flow (${\bm F}_i^F$ and ${\bm M}_i^F$) and the interparticle contact (${\bm F}_i^C$ and ${\bm M}_i^C$). In this work, the dominant fluid force/torque is the Stokes drag given by ${\bm F}^{drag} = -3\pi \mu d_p \left(\bm v - \bm u \right)f$ and ${\bm M}^{drag} = -\pi \mu d_p^3 \left({\bm \Omega} - \frac{1}{2} {\bm \omega} \right)$, where ${\bm u}$ and ${\bm \omega}$ are velocity and vorticity of the fluid, \textcolor{black}{$\mu$ is the fluid viscosity, and $d_p$ is the particle diameter. Each particle in the flow is surrounded by other particles, the presence of surrounding particles will inﬂuence the drag force for any given particle.} The friction factor $f$, given by \citet{di1994}, is used to correct for the crowding of particles. \textcolor{black}{It plays a similar role as the mobility matrix used in Stokesian dynamics for calculating the hydrodynamic drag experienced by a particle inside an agglomerate \citep{BonaJFM2014, SetoPRE2011, VanniLangmuir2011}. For particle Reynolds number in the range $0.01$ to $10^4$, $f$ can be written as}
\begin{equation}
  \label{eq:di}
  \textcolor{black}{
  f = (1-\phi)^{1-\zeta},\quad \zeta = 3.7 - 0.65 \exp\left[-\frac{1}{2}\left(1.5 - \ln {\rm Re}_p \right)^2 \right].}
\end{equation}
{\textcolor{black}{Here, $\phi$ is the local particle volume fraction and  ${\rm Re}_p$ is the particle Reynolds number, which is defined as ${\rm Re}_p=d_p |\bm v - \bm u|/\nu$.}} In addition to the Stokes drag, we also include the Saffman and Magnus lift forces in ${\bm F}^F_i$ \citep{SaffmanJFM1965, RubinowJFM1961}. {\textcolor{black}{Added mass force is neglected here, since the current work considers small and heavy particles.}

Two approaching particles interact with each other through the fluid squeeze-film between them. Such near contact interaction significantly reduces the approach velocity and further influences the collision and agglomeration process. In this work, a viscous damping force derived from the classical lubrication theory is included, given by
\begin{equation}
  \label{eq:lubrication}
  F_l = -\frac{3\pi\mu r_p^2}{2h}\frac{{\rm d}h}{{\rm d}t}.
\end{equation}
\textcolor{black}{$F_l$ is initiated at a surface separation distance $h = h_{max}\!=\! 0.01r_p$ and a minimum value of $h$, $h_{min}\!=\! 2\times10^{-4}r_p$, is set at the instant of particle contact according to experiments \citep{Marshall2011,YangPOF2006}.
The maximum value $h_{max}= 0.01r_p$ is selected such that the particles are close enough that the lubrication theory is valid. The value of $h_{max}$ is assigned according to previous work on particle-wall collision \citep{DavisJFM1986, Marshall2011}, in which simulation results yield a good ﬁt to the experimental data for restitution coefﬁcient. The minimum separation distances $h_{min}$ is set to avoid singularity. It is normally accepted that the ﬂuid density and viscosity can increase signiﬁcantly at small value of $h$, making the ﬂuid within the contact region behave in a more “solidlike” manner and limiting the value of $h$. Surface roughness will also impose a lower limit on the value of $h$ \citep{BarnockyPOF1988}. The contact mechanics are then activated when $h<h_{min}$. Setting a small gap between contacting particles has been widely adopted in contact theories (see \citet{Israelachvili2011} and references therein). The hydrodynamic force is then neglected when the two particles are in contact with each other since the contacting forces are normally much larger than the hydrodynamic force.}

When two particles $i$ and $j$ are in contact at $t_0$, the normal force $F^N$, the sliding friction $F^S$, the twisting torque $M^T$, and the rolling torque $M^R$ acting on particle $i$ from particle $j$ are expressed as
  \begin{subequations}
      \label{eq:dem}
    \begin{align}
  F_{ij}^N & \!=\!F_{ij}^{NE}\!+\!F_{ij}^{ND} \!=\!-4F_C\left(\hat{a}^3_{ij} \!-\! \hat{a}_{ij}^{3/2} \right) \!-\! \eta_N \bm{v}_{ij}\cdot \bm{n}_{ij}, \label{eq:dem_a} \\
  F_{ij}^S &\!=\! -\mathrm{min}\left[ k_T\int_{t_0}^t \bm{v}_{ij}(\tau)\cdot \bm{\xi}_S \mathrm{d}\tau \!+\!\eta_T\bm{v}_{ij}\cdot \bm{\xi}_S,\              F_{ij,crit}^S \right],        \label{eq:dem_b} \\
   M_{ij}^T &\!=\! -\mathrm{min}\left[ \frac{k_Ta^2}{2}\int_{t_0}^t \bm{\Omega}_{ij}^T(\tau)\cdot \bm{n}_{ij} \mathrm{d}\tau \!+\! \frac{\eta_Ta^2}{2}\bm{\Omega}_{ij}^T\cdot \bm{n}_{ij},\ M_{ij,crit}^T \right], \label{eq:dem_c} \\
  M_{ij}^R &\!=\! -\mathrm{min}\left[ 4F_C\hat{a}_{ij}^{3/2}\int_{t_0}^t \bm{v}_{ij}^L(\tau)\cdot \bm{t}_R \mathrm{d}\tau \!+\!\eta_R \bm{v}_{ij}^L\cdot \bm{t}_R,\ M_{ij,crit}^R \right]. \label{eq:dem_d}
\end{align}
\end{subequations}
The normal force $F_{ij}^N$ contains an elastic term $F_{ij}^{NE}$ derived from the JKR (Johnson-Kendall-Roberts) contact theory and a damping term $F_{ij}^{ND}$, which is proportional to the rate of deformation. $F^{NE}$ combines the effects of van der Waals attraction and the elastic deformation and its scale is set by the critical pull-off force, $F_C = 3\pi R_{ij}\gamma$, where $R_{ij} = (r_{p,i}^{-1} +r_{p,j}^{-1})^{-1}$ is the reduced particle radius and $\gamma$ is the surface energy density of the particle. The surface energy density $\gamma$ is deﬁned as half the work required to separate two contacting surfaces per unit area.

\textcolor{black}{
The normal dissipation coefficient $\eta_N$ in Eq. (\ref{eq:dem}a) is given as $\eta_{N}=\alpha \sqrt{m^{*} k_{N}}$, where the coefficient $\alpha$ is a function of a prescribed value of coefficient of restitution $e_0$ (see Marshall, 2009), $m^*=(m_i^{-1}+m_j^{-1})^{-1}$ is the effective mass of the two colliding particles, and the normal elastic stiffness $k_N$ is expressed as $k_N=(4/3) E_{ij} a_{ij}$. The tangential stiffness $k_T$ is expressed as $k_T=8G_{ij}a_{ij}$ and the effective elastic modulus $E_{ij}$ and shear modulus $G_{ij}$ are functions of particle’s Young’s modulus $E_i$ and Poisson ratio $\sigma_i$,
}
\begin{equation}
  \label{eq:delta_eij}
  \textcolor{black}{
\frac{1}{E_{i j}}=\frac{1-\sigma_{i}^{2}}{E_{i}}+\frac{1-\sigma_{j}^{2}}{E_{j}}, \quad \frac{1}{G_{i j}}=\frac{2-\sigma_{i}}{G_{i}}+\frac{2-\sigma_{j}}{G_{j}}
}
\end{equation}
\textcolor{black}{
where $G_i=E_i/(2(1+\sigma_i))$ is the particle’s shear modulus. The radius of contact area $a_{ij}$ is related to the value at the zero-load equilibrium state $a_{ij,0}$ through $a_{ij}=\hat{a}_{ij}a_{ij,0}$, where $a_{ij,0}$ is given as $a_{ij,0}=(9\pi \gamma R_{ij}^2/E_{ij})^{1/3}$ and $\hat{a}_{ij}$ is calculated inversely from the particle overlap, $\delta$, through \citep{JKR, ChokshiAJ1993, MarshallJCP2009}
}

\begin{equation}
  \label{eq:delta_a}
\textcolor{black}{
  \frac{\delta}{\delta_C} = 6^{\frac{1}{3}}\left[2(\hat{a}_{ij})^2 - \frac{4}{3}(\hat{a}_{ij})^{\frac{1}{2}} \right],
  }
\end{equation}
\textcolor{black}{
where $\delta_C$ is the critical overlap and is given by $\delta_{C}=a_{i j, 0}^{2} /\left(2(6)^{\frac{1}{3}} R_{i j}\right)$. The contact between the particles is built up when the overlap $\delta>0$ and is broken when $\delta<-\delta_C$. For the tangential dissipation coefﬁcient $\eta_T$ in Eqs. (\ref{eq:dem}b) and (\ref{eq:dem}c), we simply set $\eta_T=\eta_N$ \citep{TsujiPT1992}. The rolling viscous damping coefﬁcient $\eta_R$ in Eq. (\ref{eq:dem}d) is a function of coefficient of restitution $e_0$, normal elastic force $F_{ij}^{NE}$ and the effective mass of the two colliding particles $m^*$. For details, see \citet{MarshallJCP2009}.}

The sliding friction $F^S$, twisting torque $M^T$, and rolling torque $M^R$ (Eq. (\ref{eq:dem_b}) - (\ref{eq:dem_d})) are all calculated based on spring-dashpot-slider models, where $\bm{v}_{ij}\cdot\bm{\xi}_S$, $\bm{\Omega}_{ij}^T$, and $\bm{v}_{ij}^L$ are the relative sliding, twisting, and rolling velocities. When these resistances reach their critical limits, namely $F_{ij,crit}^S$, $M_{ij,crit}^T$ and $M_{ij,crit}^R$, irreversible relative sliding, twisting and rolling motions will take place between a particle and its neighboring particle. The critical limits are expressed as \citep{MarshallJCP2009}:
\begin{subequations}
    \label{eqcrit}
\begin{align}
  F_{ij,crit}^S &= \mu_{S} F_C \left|4\left(\hat{a}_{ij}^3 - \hat{a}^{3/2}_{ij}\right) + 2 \right|, \label{eqcrit_a} \\
  M_{ij,crit}^T &= \frac{3\pi a_{ij} F_{ij,crit}^S} {16}, \label{eqcrit_b}\\
  M_{ij,crit}^R &= 4F_C\hat{a}_{ij}^{3/2} \theta_{crit}R_{ij}. \label{eqcrit_c}
\end{align}
\end{subequations}
Here $\mu_S (= 0.3)$ is the friction coefficient and $\theta_{crit} (= 0.01)$ is the critical rolling angle. We set these values according to experimental measurements \citep{SumerJAST2008}. \textcolor{black}{The soft-sphere DEM for adhesive particles has been successfully applied to simulations of various systems, including particle-wall collisions \citep{ChenCES2019} and deposition of particles on a fiber \citep{YangPT2013} or on a plane \citep{LiuSM2015}, and agglomeration of particles in a pressure-driven duct flow \citep{LiuAICHE2020}, with a series of experimental and theoretical validations.}

\subsection{Simulation conditions}
Monodisperse particles are randomly seeded into the domain after the turbulence reaching the statistically stationary state. The statistical properties of the turbulent flow is fixed. Dimensionless flow parameters include the Taylor Reynolds number $Re_{\lambda} = 93.0$, the fluctuating velocity $u^{\prime} = 0.28$, the dissipation rate $\epsilon = 0.0105$, the kinematic viscosity $\nu = 0.001$, the Kolmogorov length $\eta = 0.0175$, the Kolmogorov time $\tau_k = 0.31$, and the large-eddy turnover time $T_e = 7.4$.
\textcolor{black}{These parameters together with typical scales and particle properties are listed in Table \ref{tab:kd} in both dimensional and dimensionless forms.}

The solid particles are assumed to be of micrometre scale so that the interparticle adhesion due to van der Waals attraction is expected to be the dominant force. Gravity is thus neglected here. One of the most important parameters governing the clustering of particles is the Kolmogorov-scale Stokes number, $\rm{St} = \tau_p/\tau_k$, where $\tau_p \!=\! m/(6\pi r_p \mu)$ is the particle response time and $\tau_{k} \!=\! (\nu /\epsilon)^{1/2}$ is the Kolmogorov time. In the classical theory of turbulent collision of nonadhesive particles, $\rm{St}$ significantly influences the value of the collision kernel.

The turbulent flow brings separate particles together to form agglomerates in the presence of adhesion. A sufficiently high collisional impact velocity between particles, on the other hand, gives rise to the breakage of agglomerates (collision-induced breakage, see figure \ref{fig_trj}(\textit{a})). The adhesion parameter ${\rm Ad} = \gamma/(\rho_p u^{\prime 2}r_p)$, defined as the ratio of interparticle adhesion to particle's kinetic energy, is normally used to quantify the adhesion effect \citep{LiJAS2007, Marshall2014}.
The surface energy density $\gamma$ is determined according to experimental measurements \citep{SumerJAST2008, krijt2013energy} or calculated from the Hamaker coefficients of the materials \citep{Marshall2014}. For two colliding particles, a modified adhesion number ${\rm Ad}_n = \gamma/(\rho_p v_n^2 r_p)$, which is defined based on normal impact velocity $v_n$, is often used to predict the post-collision behavior.
\textcolor{black}{
The determination of ${\rm Ad}_n$ requires the information of the normal impact velocity $v_n$, which is usually obtained from the post-processing of the simulations. One can also adopt analytical expressions to model $v_n$ (see \citet{AyalaNJP2008,PanJFM2010}) so that the value of ${\rm Ad}_n$ can be estimated before the simulations.} $\rm{Ad}$ ($\rm{Ad}_n$) has been successfully used to estimate the critical sticking velocity of two colliding particles \citep{ChenPT2015}, agglomeration efficiency of particles in turbulence \citep{ChenPRF2019}, the aerosol capture efficiency during fibre filtrations \citep{YangPT2013, ChenPRE2016}, and the packing structure of adhesive particles \citep{LiuSM2015, LiuSM2017}. In this work, we systematically vary $\rm{Ad}$ ($\rm{Ad}_n$) to show the effect of adhesion on the collision-induced breakage.

\begin{table}
  \begin{center}
\def~{\hphantom{0}}
  \begin{tabular}{lcc}
      Parameters      & Physical value  &   Dimensionless value \\[3pt]
      Typical scales   &      &     \\
      Length,$L_0$   & 0.001 m  & 1 \\
      Velocity,$U_0$   & 10 m/s   & 1 \\
      Time,$T_0$   & $10^{-4}$ s   & 1 \\
      Mass,$M_0$   & $10^{-9}$ kg   & 1 \\
         &   &   \\
      Fluid properties   &   &   \\
      Dynamic viscosity, $\mu$ & $10^{-5} \rm{Pa \cdot s}$ & -  \\
      Kinematic viscosity, $\nu$ &$10^{-5} \rm{m^2/s}$ & 0.001  \\
      Taylor Reynolds number,$Re_{\lambda}$ & - & 93.0 \\
      Fluctuating velocity, $u^{\prime}$ & 2.8 m/s & 0.28 \\
      Dissipation rate, $\epsilon$ & $1.05\times 10^4\ \rm{m^2/s^3}$  & 0.0105 \\
      Kolmogorov length, $\eta$  & $1.75\times 10^{-5}$ m & 0.0175 \\
Kolmogorov time, $\tau_k$  & $3.1\times 10^{-5}$ s & 0.31 \\
Large-eddy turnover time, $T_e$ & $7.4 \times 10^{-4}$ s &  7.4   \\
   &   &   \\
Particle properties   &   &   \\
Particle radius, $r_P$ & $5.0 \sim 12.5\ \rm{\mu m}$ & $0.005 \sim 0.0125$ \\
Particle density,$\rho_P$  & $10 \sim 320\ \rm{kg/m^3}$ &  $10 \sim 32$   \\
Surface energy, $\gamma$ & $0.01 \sim 5 \rm{J/m^2}$  &  $0.1 \sim 50$\\

  \end{tabular}
  \caption{Physical and dimensionless values of the parameters in the simulation.}
  \label{tab:kd}
  \end{center}
\end{table}

\subsection{Identification of collision, rebound and breakage events}
\label{sec2_4}
The DNS-DEM computational framework is designed with multiple-time steps  \citep{LiJAS2007, MarshallJCP2009}. \textcolor{black}{The flow field is updated with a dimensionless fluid time step ${\rm d}t_F = 0.005$, which ensures a sufficiently small Courant number. A dimensionless particle convective time step ${\rm d}t_P = 2.5\times 10^{-4}$ is adopted to update the force, velocity, and position of particles that do not collide with other particles. Such a small ${\rm d}t_p$ ensures that the distance each particle travels during a time step is only a small fraction of the particle radius so that any possible collision events can be captured. Once a particle collides with other particles during the particle time step, we then recover its information (i.e., its force, velocity, and position) to the start of the current particle time step and instead advect it with a dimensionless collision time step ${\rm d}t_C = 6.25\times10^{-6}$. The value of ${\rm d}t_C$ is small enough to resolve the rapid variation of the deformation within the contact region between touching particles (see figure \ref{fig_trj}(\textit{b})) \citep{MarshallJCP2009}.} All processes, including particle agglomeration, breakage and rearrangement of agglomerates, therefore are automatically accounted for.

Figure \ref{fig_trj}(\textit{a}) presents a typical collision-induced breakage event from the DNS-DEM simulation, where a doublet containing particles 1 (P1) and 2 (P2) collides with a third particle (P3) and then breaks into two singlets. The evolutions of interparticle overlap（scaled by the particle radius $r_p$）between P1 and P2 and that between P2 and P3 are shown in figure \ref{fig_trj}(\textit{b}). The vertical dashed lines, from left to right, mark the moment when the contact between P2 and P3 is formed, the bond between P2 and P3 and that between P1 and P2 break. The contact duration $\tau$ of each bond thus can be calculated. For instance,  $\tau_{23}$ in figure \ref{fig_trj}(\textit{b}) indicates the contact duration between P2 and P3.

To accurately interpret the breakage mechanism and formulate the breakage rate of agglomerates in turbulence, it is of crucial importance to identify various events in the simulation, including sticking of particles upon collision, rebound, collision-induced breakage and shear-induced breakage of agglomerates. We determine all these events according to the following criterion:
\begin{enumerate}[(a)]
    \item If the contact duration $\tau$ between two colliding particles is smaller than a critical value $\tau_C$, we regard it as a {\it rebound event}. In this case, there is no agglomerate formed by these two colliding particles. Rebound event normally happens when the collisional velocity is large \citep{DongPT2018, FangJAS2019}.

    \item If the bond between two colliding particles does not break within $\tau_C$, we regard it as a {\it sticking collision}. An agglomerate is then formed (or grows in size) upon the collision.

    \item When a breakage of a certain bond, whose contact duration is larger than $\tau_C$, leads to the fragmentation of an agglomerate, we regard it as a {\it breakage event}. For each breakage case, two different breakage mechanisms are further identified: \textcolor{black}{If the broken agglomerate is collided by other particles right before its breakage, we consider the breakage event as a {\it collision-induced breakage}. Otherwise, the breakage event is regarded as {\it shear-induced breakage}.}
\end{enumerate}

\begin{figure}
  \centerline{\includegraphics[width=13.8 cm]{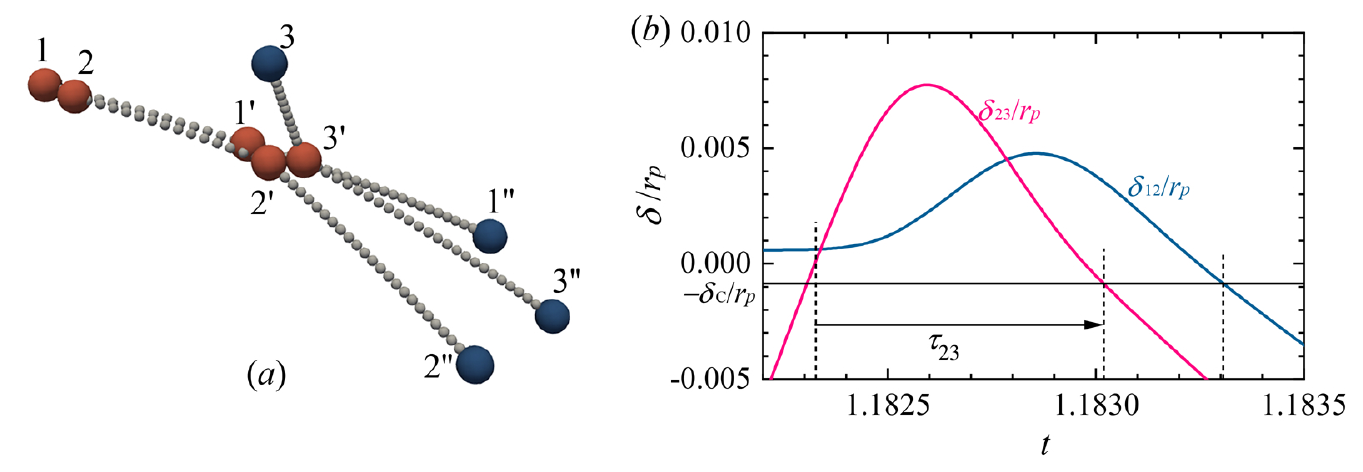}}
  \caption{(\textit{a})Trajectories of an agglomerate (doublet) and a particle from DNS-DEM simulation. $1$, $2$, and $3$ are initial positions of the particles; $1'$, $2'$, and $3'$ are corresponding particles at the collision moment; $1''$, $2''$, and $3''$ are corresponding particles at the end of trajectories. 82000 collision time steps are used to resolve the process in (\textit{a}), and the position of the particles at each 2000 time steps is presented by a grey sphere. (\textit{b}) Evolution of the interparticle overlap, where the contacting bond between particle 2 and 3 are formed at $\delta_{23} = 0$ (indicated by the vertical dashed line on the left side) and the bonds between particle 2 and 3 and particle 1 and 2 break at $\delta_{23} = -\delta_C$ and $\delta_{12} = -\delta_C$, indicated by the vertical dashed lines in the middle and on the right side, respectively.}
\label{fig_trj}
\end{figure}

To determine the value of $\tau_C$, we plot the probability distribution of the contact duration $\tau$ for the interparticle bonds in two typical cases in double logarithmic coordinates (see figure \ref{fig_pdf_tau}). There is an obvious scale separation between the contact duration in rebound events and breakage events. In the current work, the critical value $\tau_C = 0.005$ (indicated by the vertical dashed line) was chosen to separate the rebound events ($\tau < \tau_C$) and the breakage events ($\tau > \tau_C$).
The following quantities thus can be recorded in each simulation run: the number of collisions $N_C$, the number of sticking events $N_S$, rebound events $N_R$, and breakage events $N_B$.



\begin{figure}
  \centerline{\includegraphics[width=13.8 cm]{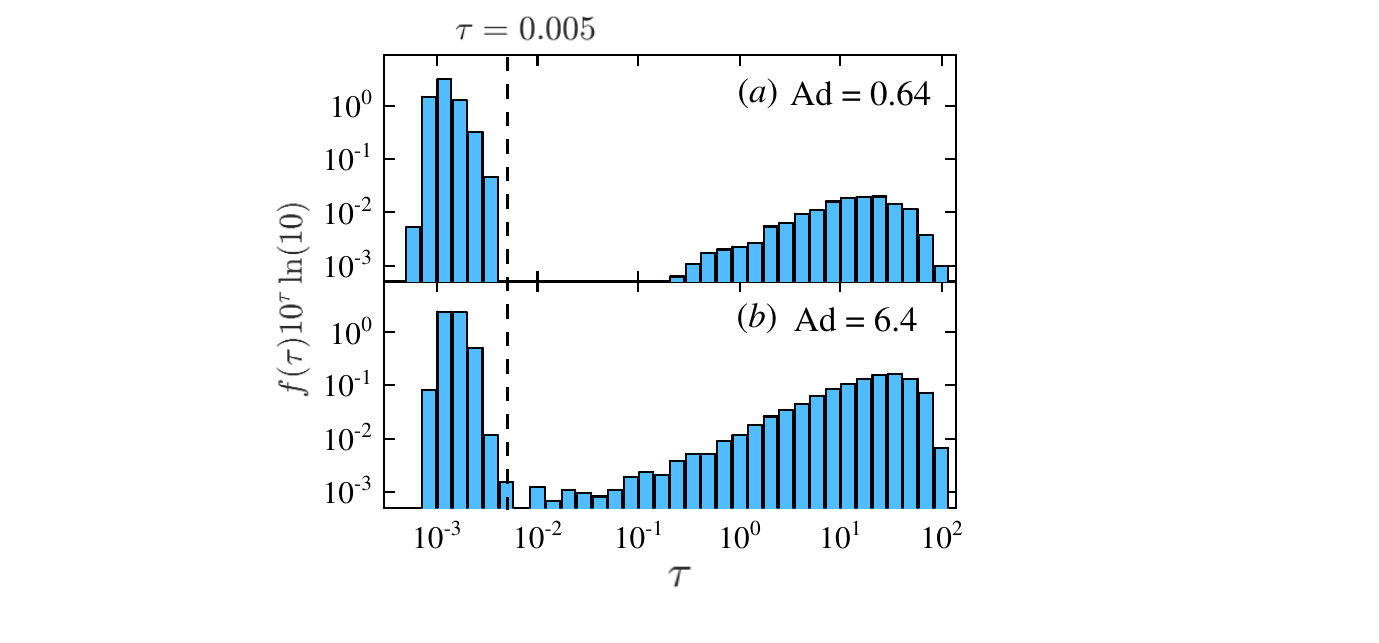}}
  \caption{Scaled probability distribution of the contact duration $\tau$ for the interparticle bonds in two typical cases with $\rm{St} = 5.8$ and (\textit{a}) $\rm{Ad} = 0.64$ and (\textit{b}) $\rm{Ad} = 6.4$. The vertical dashed line indicates the critical value $\tau = \tau_C = 0.005$, which seperates the rebound events ($\tau < \tau_C$) and the breakage events ($\tau > \tau_C$).}
\label{fig_pdf_tau}
\end{figure}

\section{Results}

\subsection{Effect of adhesion on breakage}
In figure \ref{fig_nevent1}(\textit{a}) - (\textit{c}), we show the temperal evolution of the number of overall collisions $N_C$, the number of sticking collisions $N_S$, rebound events $N_R$, and breakage events $N_B$ for $\rm{St} = 5.8$ and three different values of adhesion parameter $\rm{Ad}_n$, which is defined as
\begin{equation}
  \label{eq_adn}
  {\rm Ad}_n = \frac{\gamma}{\rho_p \bar{v}_{n}^2 r_p},
\end{equation}
where $\bar{v}_{n} = \sqrt{\left\langle v_n^{2}\right\rangle}$ is the square root of the average value of $v_n^{2}$ over all collision events. \textcolor{black}{The particles are considered to have collided at the minimum separation distance $h = h_{min}  = 2 \times 10^{-4} r_p$ and the impact velocity $v_n$ is calculated for each collision events at this moment. The values of $v_n$ are different for different collision events and $\bar{v}_{n}$ here can be regarded as an effective value to measure the kinetic energy of colliding particles.} When the adhesion is extremely weak ($\rm{Ad}_n = 0.73$), $N_C$ increases linearly with time. It indicates that the collision kernel $\Gamma$ almost keeps as a constant, which is consistent with previous DNS results for nonadhesive particles \citep{WangJFM2000}. $N_R$ is close to $N_C$ and both $N_S$ and $N_B$ are nearly zero. Agglomerates therefore can barely be formed given such a weak adhesion. For the case with a relatively stronger adhesion ($\rm{Ad}_n = 7.3$), agglomeration between colliding particles can be clearly observed. However, the agglomeration at this $\rm{Ad}_n$ value is still quite limited, since the sticking probability is small ($\sim 0.4$). When $\rm{Ad}_n$ further increases to $70$, adhesion plays a dominant role. As illustrated in figure \ref{fig_nevent1} (\textit{c}), $N_S \approx N_C$, implying that almost all collisions lead to the agglomeration of colliding particles. Moreover, $N_C$ no longer increases linearly with time in this case, which confirms previous results that intense agglomeration will push the system away from statistical equilibrium.

Another interesting result observed in figure \ref{fig_nevent1} is that the breakage of agglomerates is not obvious when the adhesion is either too weak or too strong. When $\rm{Ad}_n = 0.73$, the breakage is limited by the small number of bonds that can be formed upon collisions. In contrast, the contacting bond formed at $\rm{Ad}_n = 70$ is too strong to be broken by the fluid stress or the impact of a third particle. A considerable number of breakage events can only be observed at a moderate value of $\rm{Ad}_n$.

We normalize the number of sticking collisions $N_S$, rebound collisions $N_R$, and breakage events $N_B$ with the total number of collisions $N_C$ and plot them against $\rm{Ad}_n$ in figure \ref{fig_nevent2}. Three different regimes can be identified: a rebound regime with $\hat{N}_R > 95 \%$, a sticking regime with $\hat{N}_S > 95 \%$ and a transient regime between the above two regimes. The critical $\rm{Ad}_n$ values dividing the three regimes are approximately $1.5$ and $35$. Simulation results for different $\rm{St}$ collapse, implying that the possibility of occurrence of sticking, rebound, and breakage event can be well quantified by the dimensionless adhesion number $\rm{Ad}_n$.

\begin{figure}
  \centerline{\includegraphics[width=13.8 cm]{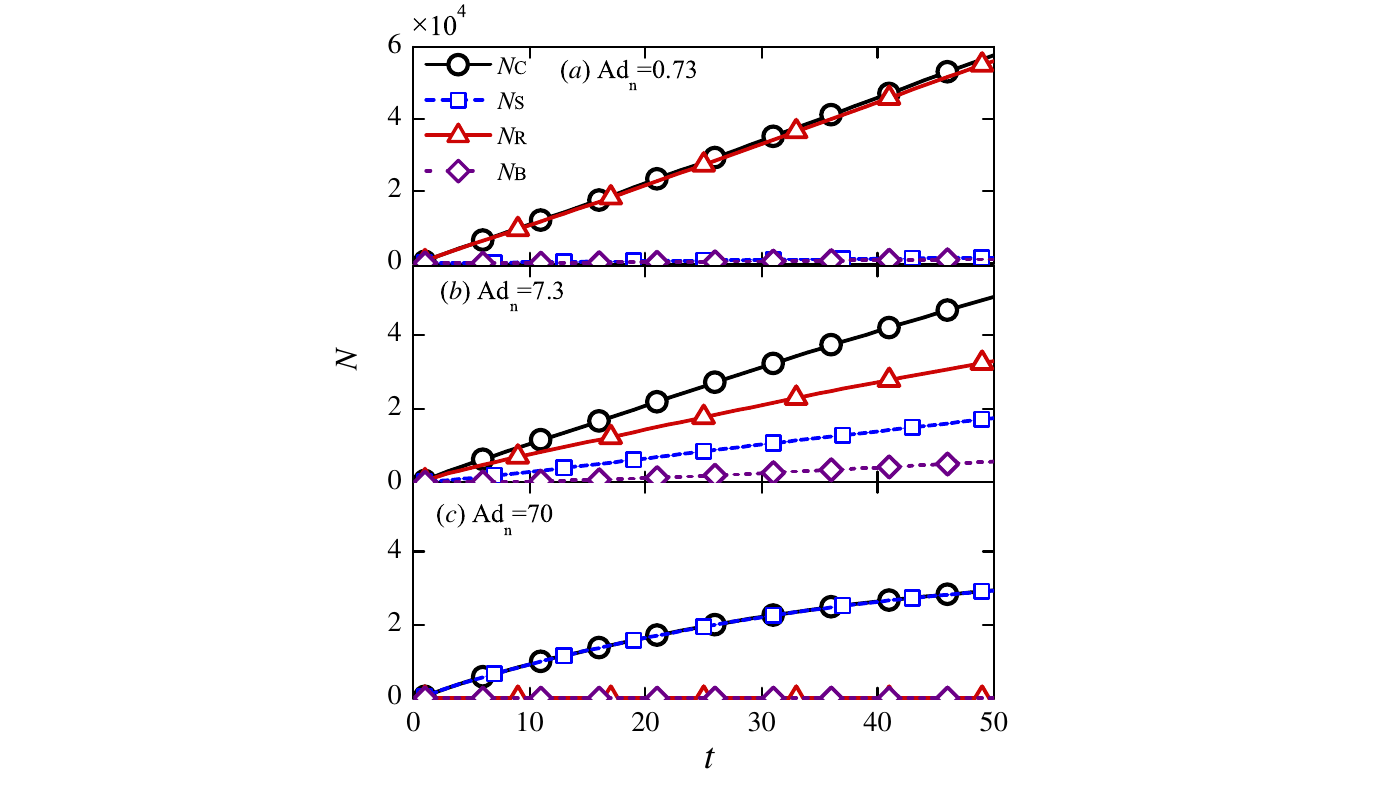}}
  \caption{Temperal evolution of the number of collisions $N_C$, the number of sticking collisions $N_S$ and rebound collisions $N_R$, and the number of breakage events $N_B$ for $\rm{St} = 5.8$ and (a) $\rm{Ad}_n = 0.73$, (b) $\rm{Ad}_n = 7.3$, and (c) $\rm{Ad}_n = 70$.}
\label{fig_nevent1}
\end{figure}

\begin{figure}
  \centerline{\includegraphics[width=13.8 cm]{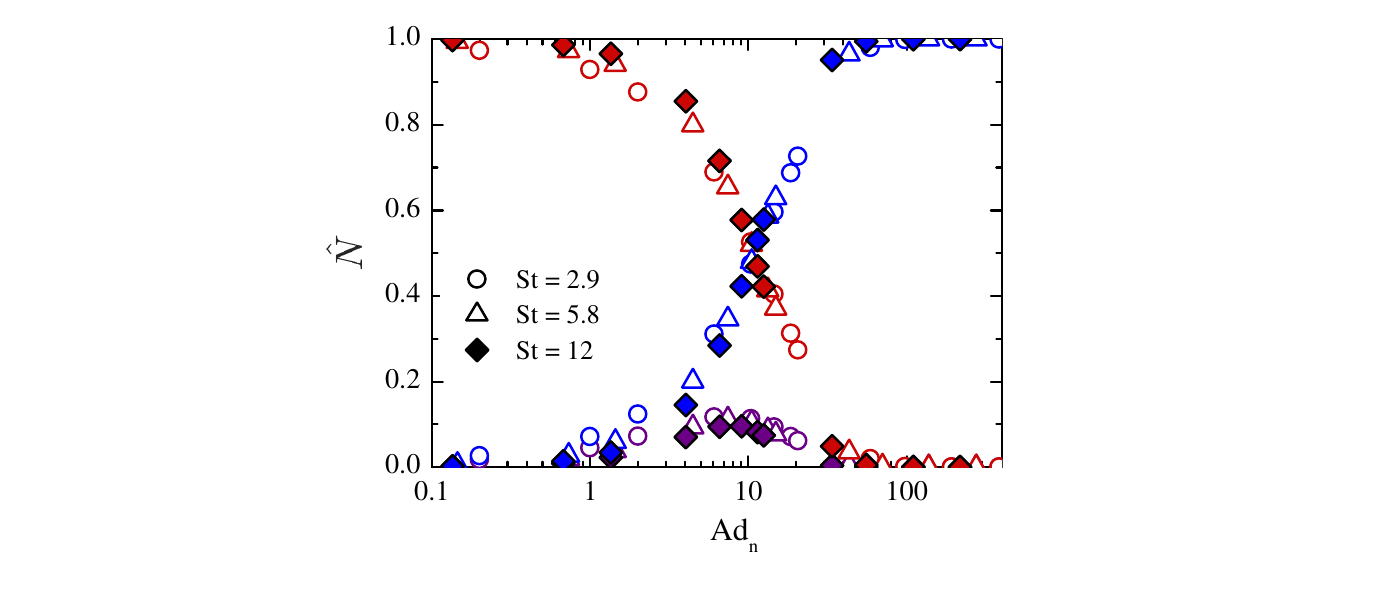}}
  \caption{Normalized number of sticking collisions $\hat{N}_S$ (blue), rebound collisions $\hat{N}_R$ (red), and breakage events $\hat{N}_B$ (purple) over the entire simulation as functions of $\rm{Ad}_n$. Results for three different Stokes numbers are shown: $\rm{St} = 2.9$ (circles), $\rm{St} = 5.8$ (triangles), and $\rm{St} = 12$ (diamonds).}
\label{fig_nevent2}
\end{figure}


\subsection{Formulation of breakage rate}
\label{sec_form}
In the current subsection, we focus on the formulation of the rate of collision-induced breakage of agglomerates. In turbulent flow laden with particles, the growth or collision-induced breakage of agglomerates results from two successive processes. First, the turbulent flow brings two initially separate agglomerates (or particles) close enough to initiate collisions. Second, the two colliding agglomerates will either merge into a large one, rebound from each other or break up into fragments.


For the first step (i.e., collision), we introduce the classic statistical model of the collision rate in particle-laden turbulence. The collision rate for agglomerates of size $i$, $\dot{n}_{C}(i)$, can be expressed as
\begin{equation}
  \label{eq_colrate}
\dot{n}_{C}(i)= \sum_{j=1}^{\infty} \Gamma(i, j) n(j)n(i),
\end{equation}
where $\Gamma(i,j)$ is the collision kernel between agglomerates of size $i$ and agglomerates of size $j$ and $n(i)$ is the the average number concentration of size group $i$. For homogenous isotropic turbulence, the collision kernel $\Gamma(i, j)$ has been modeled by \citep{ZhouJFM2001}
\begin{equation}
  \label{eq_gamma}
\Gamma(i, j)=2 \pi R_{i j}^{2}\left\langle\left|w_{r}\right|\right\rangle g\left(R_{i j}\right),
\end{equation}
where $R_{ij}$ is the radius of the effective collision spheres (ECSs) for agglomerates of size $i$ and $j$, $\left\langle\left|w_{r}\right|\right\rangle = \bar{v}_n$ is the average radial relative velocity, and $g(R_{ij})$ is the radial distribution function at the distance of contact.
\textcolor{black}{
The collision kernel $\Gamma(i,j)$ has been evaluated for non-interacting particles with different values of Stokes number in several previous studies. For monodisperse spherical particles (i.e., $i=j=1$), the collision kernel, normalized by the collision kernel for zero-inertia particles $\Gamma_0 (1,1) =(8 \pi \epsilon / 15 v)^{1 / 2}\left(2 r_{p}\right)^{3}$, increase from $1$ to $\sim10$ as $\rm{St}$ increase from $0$ to $\sim1$ and does not obviously change when $\rm{St}$ further increases \citep{SaffmanJFM1956, SundaramJFM1997, WangJFM2000, ZhouJFM2001}. In our simulation, the values of $\Gamma(1,1) / \Gamma_{0}(1,1)$ are $7.0$, $10.2$, $11.1$, and $11.0$ for $\rm{St} = 1.4, 2.9, 5.8$, and $12$, respectively. These values are quite close to the previous DNS results for non-interacting particles \citep{WangJFM2000}. The effective collision radius for an agglomerate with $i$ primary particles and that with $j$ primary particles can be calculated as $R_{ij}=R_g(i)+R_g(j)$, where $R_g(i)$ is the gyration radius for the agglomerates with $i$ primary particles \citep{JiangEST1991, flesch1999laminar, ChenPRF2019}.
}

The breakage rate due to the collisions with other particles or agglomerates can be expressed as the product of the collision rate $\dot{n}_{C}(i)$ and the fraction of collision events resulting in breakage $\Psi$ \citep{KelloggJFM2017}:
\begin{equation}
  \label{eq_col}
  f_{br}(i) = \frac{\Psi \dot{n}_{C}(i)} {n(i)} = \Psi \sum_{j=1}^{\infty} \Gamma(i, j) n(j).
\end{equation}

The fraction of breakage events $\Psi$ is defined as the ratio of the breakage number to the overall collision number. $\Psi$  should include the influence of both turbulent transport and particle scale interactions. In prior work, a critical breakage velocity $v_{b,crit}$ was introduced, assuming that agglomerate breaks when the magnitude of the normal relative velocity $v_n$ satisfies $v_n > v_{b,crit}$. The fraction of breakage events $\Psi$, therefore, can be calculated as $\Psi=\int_{v_{b, \mathrm{crit}}}^{\infty} P_C\left(v_{n}\right) d v_{n}$, with $P_C(v_n)$ being the probability density distribution of normal impact velocity \citep{KelloggJFM2017, LiuPRL2018}. Here, we introduce a new statistical framework to calculate $\Psi$ in terms of well-known impact velocity distributions $P_C(v_n)$. This formulation is expected to be more general than the previous model based on the critical breakage velocity. For collision events with impact velocity $v_n$, the fine-grained probability of breakage is recorded as $\psi(v_n)$. Thus, the distribution of velocity for breakage event is given by
\begin{equation}
  \label{eq_pdfbrk}
  P_B(v_n)=\frac{P_C(v_n) \psi\left(v_{n}\right)}{\int_0^{\infty} P_C(v) \psi\left(v\right) \mathrm{d} v},
\end{equation}
where the denominator is the normalization coefficient. $\psi\left(v\right)$ can be regarded as a transfer function, which relates the probability distribution of breakage to the impact velocity distribution.

For particles with a given adhesion value, $\psi(v_n)$ is expected to be zero as $v_n$ tends to zero (sticking regime) and rises to unity as $v_n$ increases, given that all colliding agglomerates will break when the impact velocity is sufficiently large. Knowing the value of $\psi\left(v_n\right)$, one can directly obtain the fraction of breakage $\Psi$ through
\begin{equation}
  \label{eq_Psi}
  \Psi = \int_0^{\infty} P_C(v_n) \psi\left(v_n\right) \mathrm{d} v_n.
\end{equation}
Substituting Eq. (\ref{eq_Psi}) into Eq. (\ref{eq_col}) further gives the breakage rate.


To validate the statistical framework above and \textcolor{black}{to give} a specification of the transfer function $\psi(v_n)$, we obtain the statistics of doublet breakage from DNS-DEM simulation and compare them with the theoretical descriptions in Eq.(\ref{eq_col}). The breakage of doublets has been widely adopted as the prototype of agglomerates that break into two fragments. For doublets, the breakage rate in (\ref{eq_col}) reduces to
\begin{equation}
  \label{eq_col_bdl}
  f_{br}(2) = \frac{\Psi \dot{n}_{C}(2)}{n(2)} = \Psi \sum_{j=1}^{\infty} \Gamma(2, j) n(j).
\end{equation}
At the early stage of agglomeration, most particles remain as singlets \citep{LiuPRL2018, ChenPRF2019}, the equation above can be further simplified as
\begin{equation}
  \label{eq_col_bdl2}
  f_{br}(2) \approx \Psi \Gamma(1, 2) n(1)
  = n(1)S_{12} \Psi \Gamma(1,1).
\end{equation}
On the right-hand side of the equation, we relate the singlet-doublet collision kernel $\Gamma(1,2)$ to singlet-singlet kernel through $\Gamma(1,2) = S_{12}\Gamma(1,1)$, where the constant $S_{12}$ is the correction for collisional cross section areas for singlet-doublet collisions. $\Gamma(1,1)$ for particles with different $\rm{St}$ values has been well modelled from the ghost particle approach. Although the expression in (\ref{eq_col_bdl2}) only gives low-order statistics for the breakage of doublets, it provides valuable insights: the breakage rate scale linearly to the number concentration and the effect of turbulent transport are included in both  $\Gamma(1,1)$ and the breakage fraction $\Psi$; contacting interactions affects the breakage rate by changing $\Psi$ through the transfer function $\psi(v_n)$ in (\ref{eq_Psi}).

\textcolor{black}
{In order to obtain the transfer function $\psi(v_n)$, we track all the collision events in the simulation and record whether the collision leads to the breakage of the agglomerate according to the criterion in Sec. \ref{sec2_4}. The probability distribution function of the impact velocity $P_C(v_n)$ for singlet-doublet collision events are then measured at different $\rm{St}$ and $\rm{Ad}$ values (as shown in figure \ref{fig_pdf_vn}(\textit{a}) - (\textit{c})). For the cases with weak adhesion ($\rm{Ad} = 0.64$), most particles remain as singlets and the number of singlet-doublet collision events that can be observed within a large-eddy turnover time in quite limited. We thus run three simulations with different initial random positions of particles to obtain more collision events. It ensures a good statistic on the collision velocity for singlet-doublet collision events and breakage events}. For a given value of $\rm{St}$, varying $\rm{Ad}$ does not obviously affect $P_C(v_n)$. In contrast, a strong dependence on $\rm{St}$ can be observed. For collisions that result in the breakage of a doublet, we also plot the corresponding probability distribution functions of the impact velocity, $P_B(v_n)$, in figure \ref{fig_pdf_vn}(\textit{d}) - (\textit{f}). One can easily find a strong correlation between $P_B(v_n)$ and $\rm{Ad}$. Particles with stronger adhesion tend to stick together upon collisions. The breakage events, therefore, are more likely to happen with a higher impact velocity.

\begin{figure}
  \centerline{\includegraphics[width=13.8 cm]{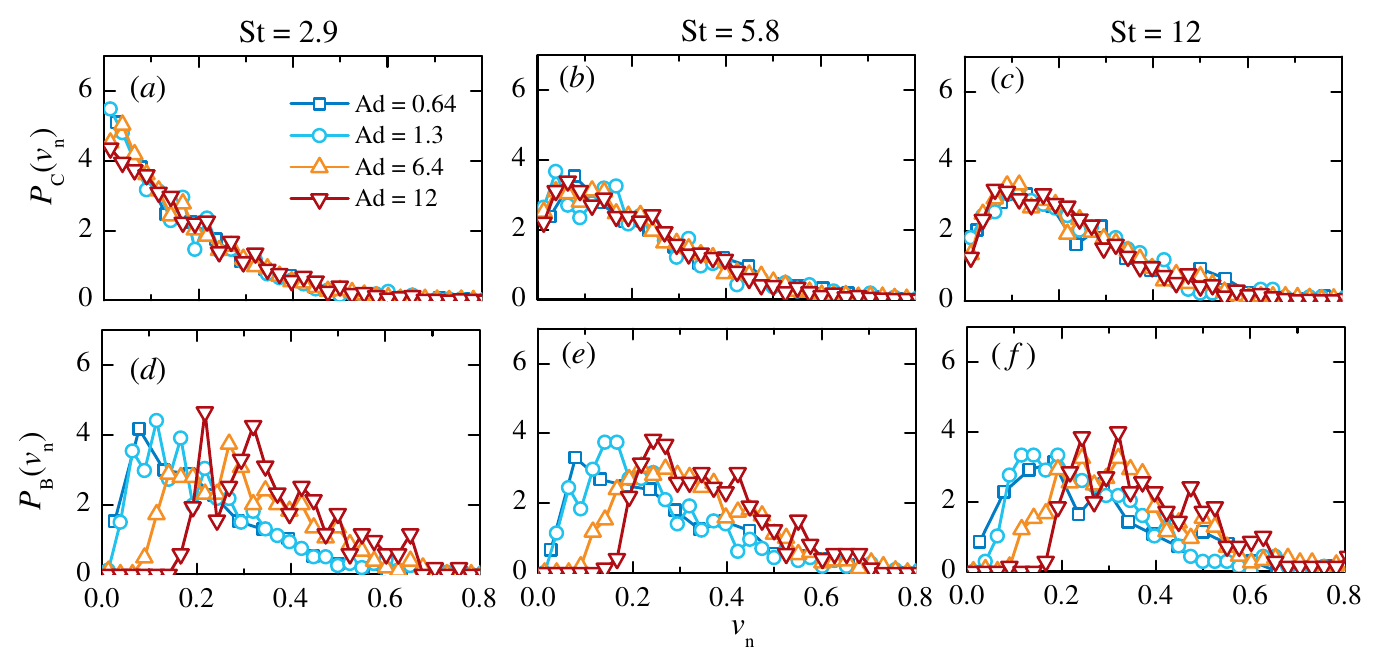}}
  \caption{Probability distribution functions of the collision velocity (normal component) $v_n$ for singlet-doublet collision events (\textit{a})-(\textit{c}) and collision-induced breakage events (\textit{d})-(\textit{f}). Statistics are made over approximately a large-eddy turnover time $t\in [15, 25]$. Different columns are results for different Stokes numbers: $\rm{St} = 2.9$ (left), $\rm{St} = 5.8$ (middle), and $\rm{St} = 12$ (right). For each Stokes number, we show results from different $\rm{Ad}$ values: $\rm{Ad} = 0.64$ (squares), $\rm{Ad} = 1.3$ (circles), $\rm{Ad} = 6.4$ (upward triangles), and $\rm{Ad} = 12$ (downward triangles). }
\label{fig_pdf_vn}
\end{figure}

We then calculate the transfer function $\psi(v_n)$ inversly from $P_C(v_n)$ and $P_B(v_n)$ according to (\ref{eq_pdfbrk}). As shown in figure \ref{fig_transformation}(\textit{a}), despite the inconsistency in $P_C(v_n)$, $\psi(v_n)$ for different $\rm{St}$ nicely collapses. In contrast, the adhesion strongly affects $\psi(v_n)$. Although, there is considerable scatter in the data at large $v_n$ due to the limited sample size of the energetic collision events, the transfer function $\psi(v_n)$ at a given $\rm{Ad}$ value is roughly linear to the collision velocity $v_n$. The results in figure \ref{fig_transformation} (\textit{a}) suggest that the transfer function may only depends on the short-range contacting interactions, whereas the effects of turbulent transport and hydrodynamic interactions are included in the probability distribution functions of the impact velocity $P_C(v_n)$. To validate the argument above, we run simulations with different particle radius (ranging from 0.0075 to 0.0125) and with/without the hydrodynamic damping force (Eq. \ref{eq:lubrication}) at a fixed ${\rm St}$ value. As seen in figure \ref{fig_transformation}(\textit{a}), the measured transfer function $\psi(v_n)$ does not show obvious dependence on the particle size and the hydrodynamic interaction, confirming that the transformation function $\psi(v_n)$ is determined by the short-range contacting interactions.

According to the results in figure \ref{fig_transformation}(\textit{a}), we propose a linear relationship between $\psi$ and $v_n$:
\begin{equation}
    \label{eq_psi_fitting}
\psi(v_n)=\left\{\begin{array}{ll}
{0,} & {\text { for } v_n < v_{C1},} \\
{\frac{1}{v_{C2} - v_{C1} } (v_n - v_{C1}),} & {\text { for } v_{C1} \leq v_n \leq v_{C2},} \\
{1,} & {\text { for } v_n > v_{C2}.}
\end{array}\right.
\end{equation}
Two typical values of collision velocity $v_{C1}$ and $v_{C2}$ are indicated by Eq. (\ref{eq_psi_fitting}). Breakage does not happen when the collision velocity between two agglomerates, $v_n$, is smaller than $v_{C1}$. On the other hand, if $v_n > v_{C2}$, the colliding doublets always break. We then fit the measured values of the transfer function $\psi(v_n)$ (linear part) using Eq. (\ref{eq_psi_fitting}) for all the cases presented in figure \ref{fig_transformation} (\textit{a}) and plot the fitting parameters $v_{C1}$ and the slope $(v_{C2} - v_{C1})^{-1}$ as a function of $\rm{Ad}$. It is seen that the fitted values of the slope for different cases center around a logarithmic curve (figure \ref{fig_transformation} (\textit{b})), which reads
\begin{equation}
\label{eq_slopad_fitting}
(v_{C2} - v_{C1})^{-1} = -2.1 \ln \left( \frac{\rm{Ad}}{13}\right).
\end{equation}
Several interesting features are indicated by Eq. (\ref{eq_slopad_fitting}). First, the slop diverges in the small adheison limit ($\rm{Ad} \to 0$), indicating that there is a critical collision velocity separating the breakage and non-breakage collisions. This is in accordance with the
theoretical model proposed by \citet{LiuPRL2018}, in which a Heaviside function $H(v - v_{b,crit})$ is proposed to transform the PDF of normal impact velocity $P_C(v_n)$ into the PDF of impact velocity for breakage events $P_B(v_n)$. We show here that such transfer function is reasonable only when the adhesive interaction is extremely weak. As $\rm{Ad}$ increases, the slope of $\psi(v_n)$ considerably decreases and there is no sharp transition between breakage and non-breakage collision velocities. Although the data points for the minimum breakage velocity $v_{C1}$ are relatively dispersed when plotted as a function of $\rm{Ad}$, a quadratic curve, $v_{C1} = a\rm{Ad}^2$ with $a = 7.4\times 10^{-4}$, can roughly describe the variation of $v_{C1}$ (see figure \ref{fig_transformation} (\textit{c}))). $v_{C1}$ diverges at large adhesion limit, implying that all collisions give rise to the growth of agglomerates when the adhesion is sufficiently strong.


\begin{figure}
  \centerline{\includegraphics[width=13.8 cm]{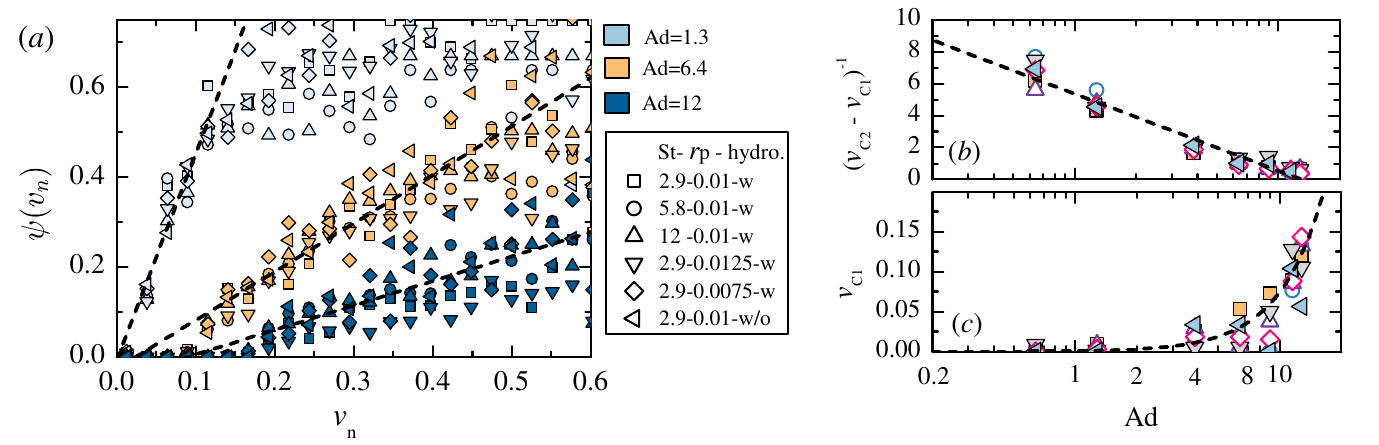}}
  \caption{(\textit{a}) Transfer function $\psi\left(v_{\mathrm{n}}\right)$ versus collision velocity $v_n$ for different Stokes numbers: $\rm{St} = 2.9$ (squares), $5.8$ (circles), and $12$ (triangles), and different $\rm{Ad}$ values: $\rm{Ad} = 1.3$ (light blue), $6.4$ (yellow), $12$ (dark blue). Results with different particle radius (ranging from 0.0075 to 0.0125) and with/without the hydrodynamic damping force (Eq. \ref{eq:lubrication}) at ${\rm St = 2.9}$ are also included. Scatters are results calculated from PDFs in figure \ref{fig_pdf_vn}, and dashed lines are linear-fittings from Eq. (\ref{eq_psi_fitting}). (\textit{b}) and (\textit{c}) Fitting parameters $(v_{C2} - v_{C1})^{-1}$ and $v_{C1}$ as functions of $\rm{Ad}$. Legends are the same as in (\textit{a})}
\label{fig_transformation}
\end{figure}

To further validate the model of the transfer function, we present an example of the model prediction for cases with $\rm{St}= 2.9$ in figure \ref{fig_psi} (\textit{a}). First, the probability distribution function of the normal collision velocity $P_C(v_n)$ is measured from the simulation with small $\rm{Ad}$ value ($1.3$). The breakage fraction $\Psi$ is then calculated by substituting Eq. (\ref{eq_psi_fitting}) and the measured $P_C(v_n)$ into Eq. (\ref{eq_Psi}). One can also adopt models of $P_C(v_n)$ obtained from simulations with non-interacting particles to estimate the breakage fraction $\Psi$ \citep{saw2014extreme, BhatnagarPRE2018, SalazarJFM2012}. Such approximation does not bring large errors since $P_C(v_n)$ is almost independent of adhesive interactions (see figure \ref{fig_pdf_vn}). The result generated from the model together with predictions for $\rm{St} = 1.4$, $5.8$ and $11.5$ is plotted as a dash line in figure \ref{fig_psi}(\textit{b}). We see that the model predictions are in accordance with DNS-DEM simulations.
\textcolor{black}
{The deviation between the model and the simulations in figure \ref{fig_psi}(\textit{b}) may result from the linear assumption of the transfer function $\psi(v_n)$ (Eq.(\ref{eq_psi_fitting})), in which a sharp transition is assumed between the linear part ($(v_n-v_{C1})/(v_{C2}-v_{C1})$) and unity. The simulation data in figure \ref{fig_transformation}(\textit{a}), in contrast, shows a much slower approach to unity, indicating that the model in Eq. (\ref{eq_psi_fitting}) overestimates $\psi(v_n)$ when $v_n \to v_{C2}$. Despite this deviation, our simplified model well captures the variation of breakage fraction $\Psi$ with adhesion $\rm{Ad}$.} Moreover, Stokes number dependence of $\Psi$ can be observed in figure \ref{fig_psi} (\textit{b}). Since the breakage fraction $\Psi$ here is calculated from a universal transfer function, the $\rm{St}$ number dependence of $\Psi$ originates from the difference in $P_C(v_n)$: the hydrodynamic damping force significantly reduces the relative approaching velocity of colliding particles with small $\rm{St}$.

\begin{figure}
  \centerline{\includegraphics[width=13.8 cm]{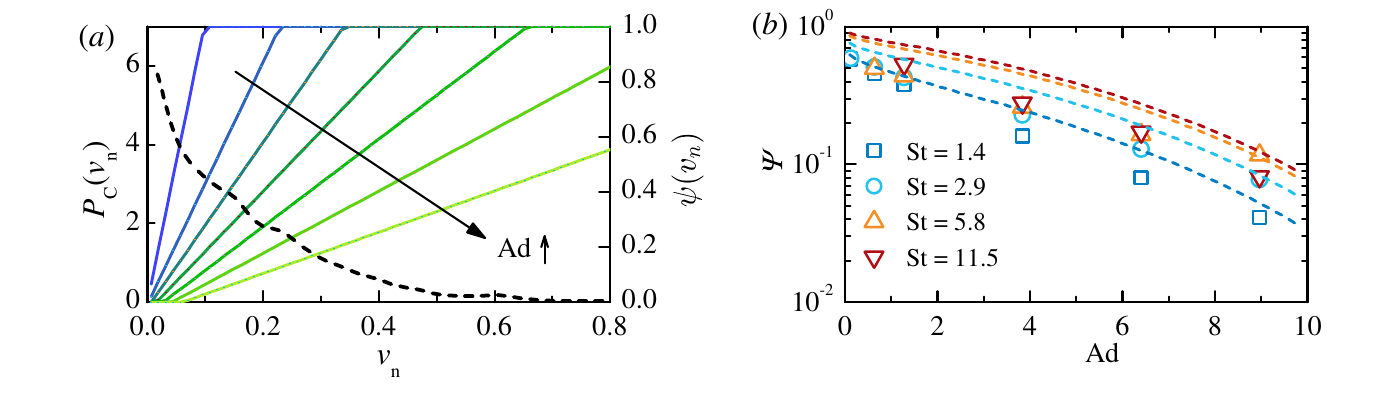}}
  \caption{(\textit{a}) Probability distribution functions $P_C(v_n)$ of the normal collision velocity for $\rm{St} = 2.9$ and $\rm{Ad} = 1.3$ (left axis) and the transfer function $\psi(v_n)$ modeled by Eq. (\ref{eq_psi_fitting}) (right axis). Color code spans from blue to yellow with increasing $\rm{Ad}$ (from $0.1$ to $10$). (\textit{b}) Fraction of collision-induced breakage of doublets $\Psi$ at different $\rm{Ad}$ values. Points are DNS-DEM results and dashed lines are results calculated from $P_C(v_n)$ and the modeled $\psi(v_n)$ (Eq.(\ref{eq_psi_fitting})).}
\label{fig_psi}
\end{figure}

The collision-induced breakage rate of the doublets $f_{b r}(2)$ is calcualted from Eq.(\ref{eq_col_bdl2}) and compared with DNS-DEM results in figure \ref{fig_fbrcompare}(\textit{a}). Quantitative agreement is observed, indicating that the analytical model well captures the effects of the particle inertia and the adhesive interaction on the breakage. Since the adhesion parameter $\rm{Ad}$ does not include the effect of particle inertia, there is considerable distinction in results for different $\rm{St}$ at the same $\rm{Ad}$. We stress again that particle inertia affects the breakage rate through its influence on the statistics of the collision velocity. One simple way to include both effects of particle inertia and the adhesion is to use the modified adhesion parameter $\rm{Ad}_{n}$ (see Eq. (\ref{eq_adn})), which scales the adhesion using $\rm{St}$-dependent avereage velocity $\bar{v}_{n} = \sqrt{\left\langle v_n^{2}\right\rangle}$. The normalized breakage rate, when plotted as a function of $\rm{Ad}_{n}$, nicely collapse onto the exponential curve (see figure \ref{fig_fbrcompare}(\textit{b})):
\begin{equation}
  \label{eq_fbr_vs_adn}
\frac{f_{br}\tau_k}{r_p^3 n(1)} = 86\exp(-0.12\rm{Ad}_n).
\end{equation}
The result indicates that $\bar{v}_{n} = \sqrt{\left\langle v_n^{2}\right\rangle}$ is an appropriate choice to scale the effect of adhesion and the collision-induced breakage rate can be well estimated once $\rm{Ad}_n$ is known. Current values of $\bar{v}_{n}$ are measured from DNS-DEM simulation.

It should be noted that the model in Eq. (\ref{eq_col_bdl2}) is valid only for early-stage agglomeration, since the transfer function is derived for singlet-doublet collisions. Both agglomerate size and structure may affect the formulation of the transfer function. Predicting the breakage rate for agglomerates with arbitrary size and structures through first principles is practically impossible. It is thus normally accepted to describe the breakage rate using an exponential or a power-law function, in which the parameters are related to agglomerate size and particle-particle interactions. Agglomerate size dependence of the breakage rate will be discussed in Sec. \ref{sec_sizeeffect}.

\begin{figure}
  \centerline{\includegraphics[width=13.8 cm]{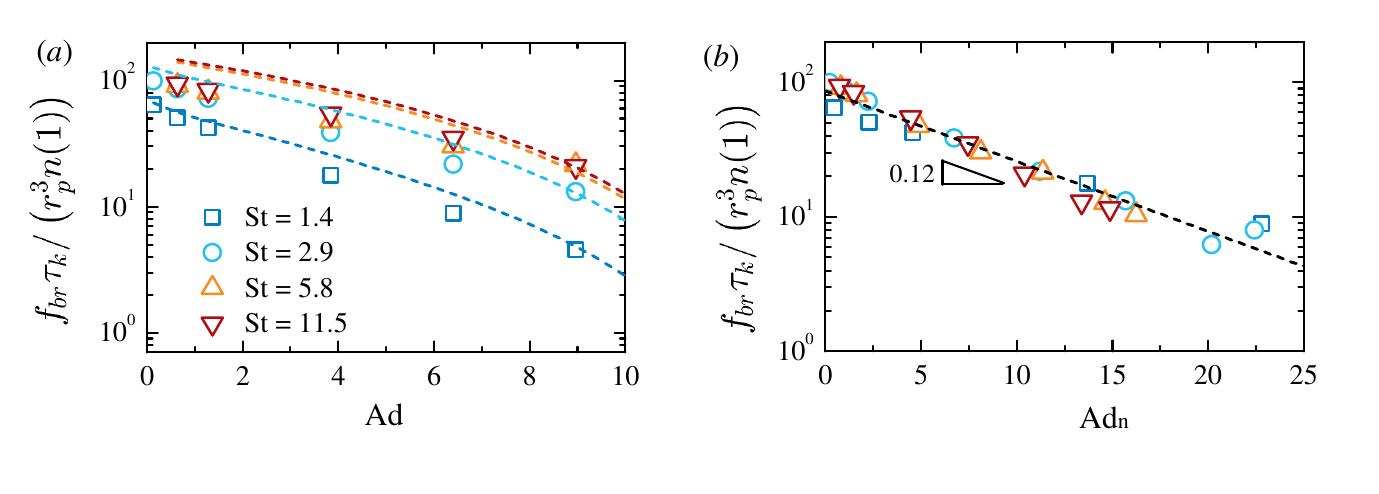}}
  \caption{(\textit{a}) Normalized breakage rate $f_{br}\tau_k/(r_p^3 n(1))$ for doublets as a function of $\rm{Ad}$. The scatters are DNS-DEM results and the dashed lines are predictions from Eq. (\ref{eq_col_bdl2}), in which the breakage fraction $\Psi$ is calculated from the pdf of normal collision velocity through $ \Psi=\int_{0}^{\infty} P_{C}(v_n) \psi(v_n) \mathrm{d} v$ (see Eq. (\ref{eq_Psi})) and the transfer function $\psi(v_n)$ is modeled by Eq. (\ref{eq_psi_fitting}). (\textit{b}) Normalized breakage rate as a funcion of $\rm{Ad_n}$. The dashed line is exponential fittings using Eq. (\ref{eq_fbr_vs_adn})}
\label{fig_fbrcompare}
\end{figure}

It is of great importance to know how the breakage rate scales with particle size $r_p$ and particle number density $n$. We measure the doublet breakage at different particle size ($r_p = 0.005 \sim 0.015$) and particle numbers ($N = 19600 \sim 40000$) for typical $\rm{St}$ and $\rm{Ad}$ values (shown in figure \ref{fig_rpnpscaling}). The results are plotted in a scaled form: $\hat{f}_{br} = f_{br}(r_p)/f_{br}(r_{p,0})$, $\hat{r}_p = r_p/r_{p,0}$ in figure \ref{fig_rpnpscaling} (\textit{a}) and $\hat{f}_{br} = f_{br}(n(1))/f_{br}(n_{m}(1))$, $\hat{n}(1) = n(1)/n_{m}(1)$ in figure \ref{fig_rpnpscaling} (\textit{b}). Here, $f_{br}(r_{p,0})$ is the doublet breakage rate for the case with $r_{p,0} = 0.01$ and $f_{br}(n_{m}(1))$ is the breakage rate for the case with the maximum value of singlet number density $n_{m}(1)$. As displayed in figure \ref{fig_rpnpscaling}, DNS-DEM results follow the power laws $\hat{f}_{br} \propto \hat{r}^2_p$ and $\hat{f}_{br} \propto \hat{n}^1(1)$ when particle size and singlet number density are varied.

The $n(1)$ dependence is easy to understand from (\ref{eq_col_bdl2}). The $\hat{r}^2_p$ scaling originates from the $r_p$ dependence of the collision kernel $\Gamma(1,1)$ in (\ref{eq_col_bdl2}). For inertial particles ($\rm{St} \gg 1$), the approaching velocity of colliding particles is decorrelated from the local fluid gradient, thus is not affected by the particle size. The $\hat{r}^2_p$ scaling enters $\Gamma(1,1)$ through the effective collision area. We note that the size scaling here is valid for particles that are smaller than or comparable to the Kolmogorov scale. It may not hold for particles that are considerably larger than the Kolmogorov scale. In the latter case, particle-resolved simulations would be needed to precisely calculate the flow around and forces on particles \citep{ErnstAM2013, WangCES2019, LiuPOF2019, PengJFM2019}.

\begin{figure}
  \centerline{\includegraphics[width=13.8 cm]{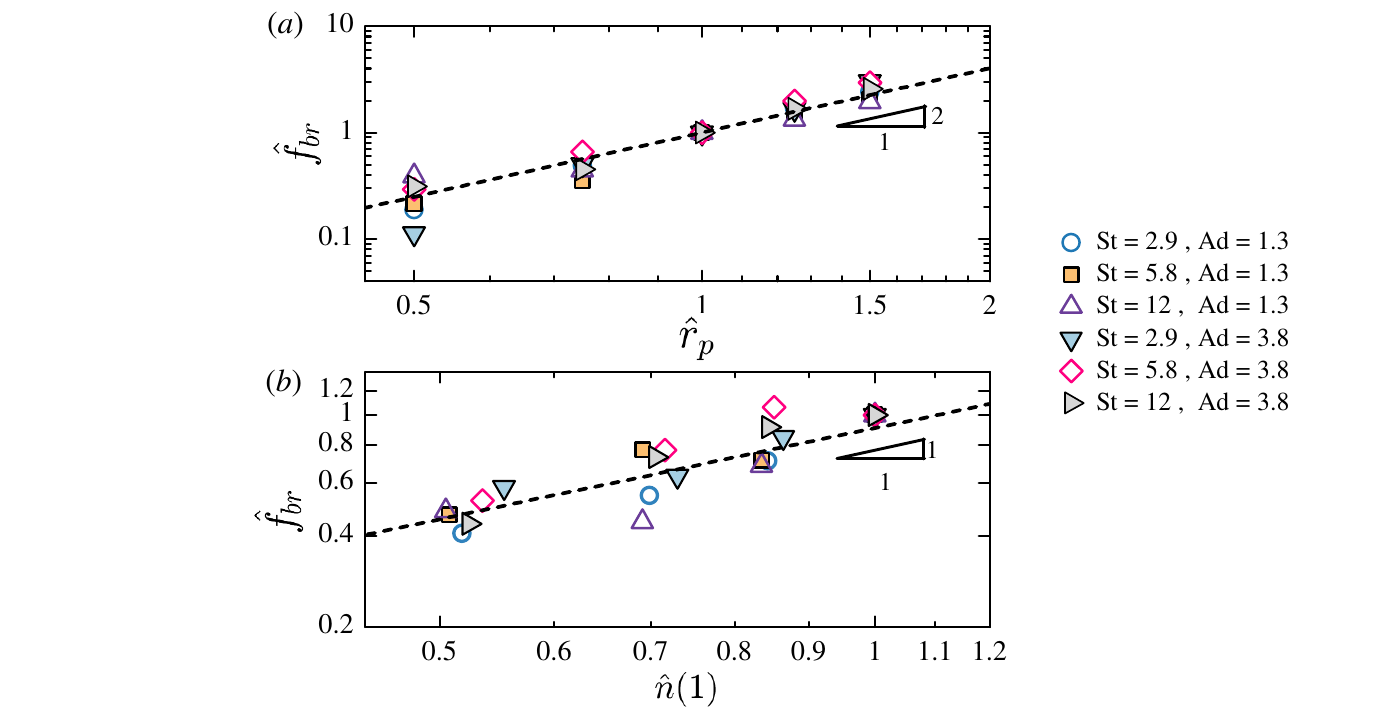}}
  \caption{Scaled breakage rate $\hat{f}_{br}$ as a function of (\textit{a}) scaled particle size $\hat{r}_p$ and (\textit{b}) scaled number density of singlet $\hat{n}(1)$ at $\rm{St} = 2.9$, $5.8$, and $12$, and $\rm{Ad} = 1.3$ and $3.8$. Dashed lines in (\textit{a}) and (\textit{b}) indicate power law  functions with exponents $2$ and $1$, respectively.}
\label{fig_rpnpscaling}
\end{figure}

\subsection{Agglomerate size dependence of the breakage rate}
\label{sec_sizeeffect}
The breakage rate of agglomerates with size $A$ are calculated from DNS-DEM simulations according to
\begin{equation}
  f_{br}(A) = \frac{N_{br}(A)}{N(A)\Delta t}
\end{equation}
where $N(A)$ is the number of agglomerates of size A averaged over the time range $t\in[30,40]$, $N_{br}(A)$ is the breakage number of agglomerates with size $A$, and $\Delta t = 10$. As shown in figure \ref{fig_sizeeffect} (\textit{a}), a stronger adhesion promotes the formation of larger agglomerates. In contrast, the number of breakage decreases with $\rm{Ad}$ (see figure \ref{fig_sizeeffect} (\textit{b})). To provide meaningful statistics, we only calculate $f_{br}(A)$ when $N_{br}(A)$ is larger than $20$. The results are normalized by the mean shear rate $G$ and plotted as a funcion of size $A$ in figure \ref{fig_sizeeffect} (\textit{c}). It is seen that the breakage rate depends linearly on the agglomerate size with the slope being a function of $\rm{Ad}_n$. Fitting the data at different $\rm{Ad}_n$ and $\rm{St}$ according to
\begin{equation}
  \label{eq_fbr_vs_A}
  \frac{f_{b r}(A)}{G}=\zeta({\rm St}, {\rm Ad}_n) \cdot A +\chi
\end{equation}
gives us the values of the slope $\zeta (\rm{St},\mathrm{Ad}_n)$. As shown in figure \ref{fig_sizeeffect}(\textit{d}), when plotted as a function of $\mathrm{Ad}_n$, $\zeta$ for different $\rm{St}$ centers around a universal curve, which is analogous to the $f_{br}$ dependence in Fig. \ref{fig_fbrcompare}({\it b}). The universal curve has an power-law form: $\zeta = 0.012\cdot \rm{Ad}_n^{-0.81}$. These results once again confirm that the modified adhesion parameter $\rm{Ad}_n$ is an appropriate choice to reflect both effects of the particle inertia and adhesive interactions on the breakage.

\begin{figure}
  \centerline{\includegraphics[width=13.8 cm]{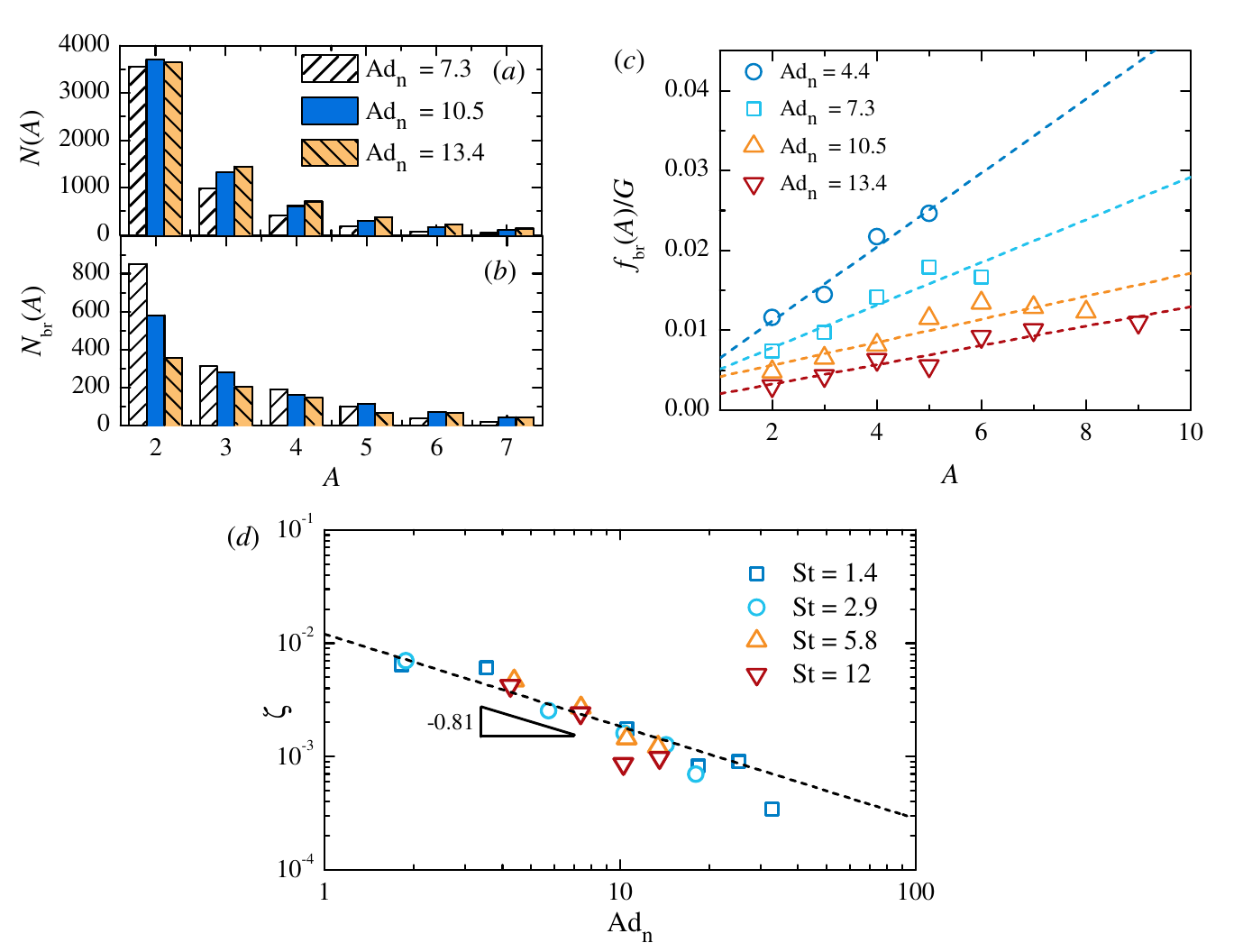}}
  \caption{(\textit{a}) Number of agglomerates of size $A$ averaged over the time range $t\in[30,40]$. (\textit{b}) Number of breakages of agglomerates of size $A$ measured within $t\in[30,40]$ for $\rm{St} = 5.8$. (\textit{c}) Breakage rate, normalized by the shear rate $G$, as a function of agglomerate size $A$. The dashed lines are linear fittings from Eq. (\ref{eq_fbr_vs_A}). (\textit{d}) The fitting values of the slope $\zeta$ for the linear relationship between $f_{br}/G$ and $A$ at different $\rm{Ad}_n$ and $\rm{St}$ values. The dashed line indicates the power function $\zeta = 0.012\cdot \rm{Ad}_n^{-0.81}$. }
\label{fig_sizeeffect}
\end{figure}

\subsection{Role of flow structure}
In this subsection, we quantify the correlation between structures of turbulence and the breakage of agglomerates with different $\rm{St}$ and $\rm{Ad}$ values. We identify the flow structures based on the second invariant of the velocity gradient tensor $\mathcal{Q}=\left(\mathcal{R}^{2}-\mathcal{S}^{2}\right) / 2$, where the strain rate tenor $\mathcal{S}=\left(\mathcal{A}+\mathcal{A}^{\mathrm{T}}\right) / 2$ and the rotation rate tensor $\mathcal{R}=\left(\mathcal{A}-\mathcal{A}^{\mathrm{T}}\right) / 2$ are symmetric and antisymmetric part of the velocity gradient tensor $\mathcal{A}=\tau_{k} \nabla \boldsymbol{u}$ (normalized by the Kolmogorov time $\tau_{k}$), respectively. Figure \ref{fig_vortex} (\textit{a}) presents the countour plots of $Q$, showing the vortex tubes with $Q > 3.3\sqrt{\langle Q^2 \rangle}$ and straining sheets with $Q < -2.5\sqrt{\langle Q^2 \rangle}$, and the corresponding 2-D slice at $y=0$. One can clearly see the red vortex tubes surrounded by blue straining sheets (vortex-strain worm-rolls), which implies that intense structures typically occur near each other \citep{PicardoPRF2019}.


We calculate the average $\mathcal{Q}$, sampled by singlet-doublet collisions, at different $\rm{St}$ and $\rm{Ad}$ values in figure \ref{fig_vortex}(\textit{b}). The results for non-interacting particles based on ghost collision approximation are also included \citep{PicardoPRF2019} (only data at $\rm{St} > 0.5$ are shown here). One can notice that as $\rm{St}$ increases from $0.5$ to $20$, $\mathcal{Q}$ increases from a negative value to zero, implying that finite-inertia particles (${\rm St}\sim 1$) tend to collide in the straining zone whereas particles with large inertia collide uniformly. According to \citet{PicardoPRF2019}, decreasing $\rm{St}$ also leads to $\mathcal{Q}$'s approach to zero and the largest absolute value of $\mathcal{Q}$ occurs at $\rm{St} \approx 0.3$. Such flow structure dependence is owing to two aspects. First, particles with finite inertia ($\rm{St} \approx 1$) tend to accumulate in straining regions outside vortices due to the centrifugal effect (known as preferential concentration). Moreover, particle inertia also increases the relative approaching velocity between particles. Such effect also prevails in straining zones \citep{PicardoPRF2019}. Here, we show that varying particle-particle contacting interactions ($\rm{Ad}$) does not obviously affect the structure dependence of collisions.

The average $\mathcal{Q}$, sampled by singlet-doublet breakage events, shows a strong dependence on $\rm{Ad}$ (figure \ref{fig_vortex}(\textit{c})). Doublets with larger $\rm{Ad}$ value are more difficult to break thus needs higher impact velocities. For particles with moderate inertia ($\rm{St} \approx 1$), violent collisions are more likely caused by particles ejected rapidly from strong vortices and happen in straining sheets (with smaller negative $\mathcal{Q}$) that envelope the vortices. As $\rm{St}$ increases, the relative velocity between colliding particles becomes less sensitive to the underlying flow, both collision events and breakage events distribute more uniformly in the flow. As shown in figure \ref{fig_vortex} (\textit{d}), the relationship between $\mathcal{Q}$ and $\rm{Ad}$ at given $\rm{St}$ can be well described by linear functions.

\begin{figure}
  \centerline{\includegraphics[width=13.8 cm]{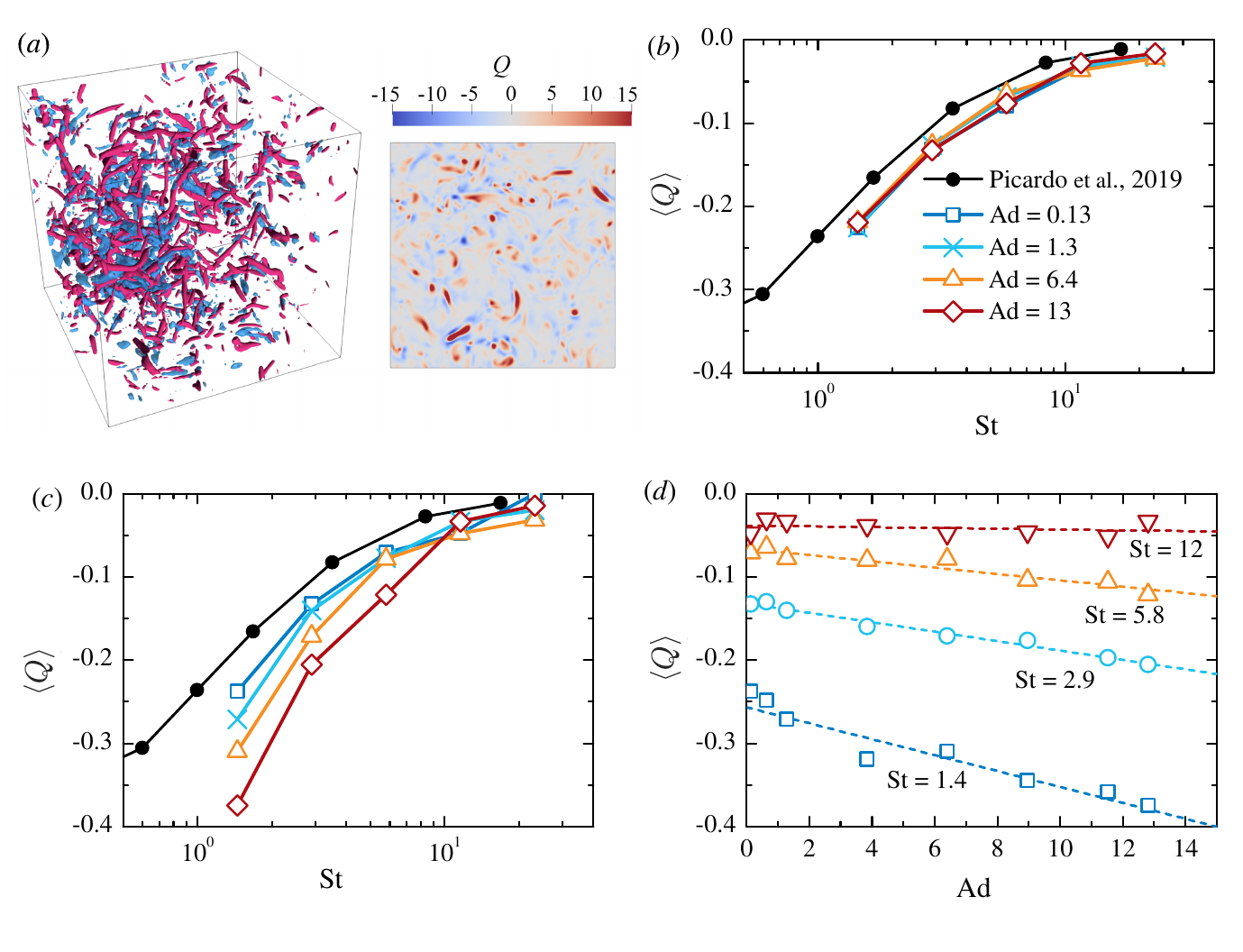}}
  \caption{(\textit{a}) Countour plot of $Q$ and the 2-D slice at $y=0$. Vortex tubes with $Q > 3.3\sqrt{\langle Q^2 \rangle}$ are colored in red and straining sheets with $Q <-2.5\sqrt{\langle Q^2 \rangle}$ are colored in blue. (\textit{b}) Average $Q$, sampled by singlet-doublet collision events and (\textit{c}) average $Q$ sampled by doublet breakage events as functions of $\rm{St}$. (\textit{d}) Average $Q$ for breakage events as a function of $\rm{Ad}$ at different $\rm{St}$ values: $\rm{St}= 1.4$ (squares), $\rm{St} = 2.9$ (circles), $\rm{St} = 5.8$ (upward triangles), and $\rm{St} = 12$ (downward triangles). The straight dashed lines are linear fittings.}
\label{fig_vortex}
\end{figure}

\section{Discussion \& Conclusions}
By means of DNS and multiple time scale DEM, we are able to resolve all the collision, rebound, and breakage events for adhesive particles in turbulence. We have shown that the collision-induced breakage rate of agglomerates can be modelled based on the statistics of the collision rate and a breakage fraction function $\Psi$. A scaling relationship of the breakage rate for doublets at the early stage is proposed, which includes the effects of particle size, turbulent transport, and particle number concentration. The fraction function $\Psi$ is further expressed as a function of the well-known distributions of impact velocity and a universal transfer function $\psi(v_n)$, which is shown to rely on particle-particle contacting interactions and is independent of particle inertia $\rm{St}$, particle size, and hydrodynamic interactions. Based on a large number of simulations, we propose an exponential function of adhesion parameter $\rm{Ad}_n$ for the breakage rate of doublets and show that the breakage rate increases linearly as the agglomerate size increases. The framework allows one to estimate the breakage rate for early-stage agglomerates of arbitrary size.

It is of great interest to compare our results with {\textit shear-induced breakage} of agglomerates, which has been extensively investigated for both isostatic loose agglomerates \citep{BonaJFM2014} and dense ones. The shear-induced breakup rate for doublets scales exponentially with a dimensionless parameter, $\mathcal{N}$ \citep{BonaJFM2014}:
\begin{equation}
  \label{eq_fbr_sh}
  f_{br}^{\rm{sh}}= \frac{k_{f}}{\tau_{k}} \exp (-\alpha \mathcal{N})
\end{equation}
where $\mathcal{N}=F_{C} /\left(6 \pi \mu r_p^{2} {G}_{\rm{eff}}\right)$, $F_{C}$ is the strength of the bond and $6 \pi \mu r_p^{2} G_{\rm{eff}}$ estimates the largest tensile stress acting on the bond by the flow field with an effective shear rate ${G}_{\rm{eff}}$. $k_{f}$ is a prefactor of order unity and $\alpha$ is fitted to be $4.8$ for $\mathcal{N} <0.5$ and $1.8$ for $\mathcal{N} > 1.5$. According to (\ref{eq_fbr_vs_adn}) and (\ref{eq_fbr_sh}), the ratio between collision-induced breakage rate $f_{br}^{co}$ and shear-induced breakage rate $f_{br}^{sh}$ can be estimated as

\begin{equation}
  \label{eq_fbr_ratio}
\frac{f_{b r}^{s h}}{f_{b r}^{c o}} \sim \phi^{-1} \frac{\exp \left(-0.25\alpha  \gamma r_{p}^{-1}  \mu^{-1} {G}_{\mathrm{eff}}^{-1}  \right)}{\exp (-0.12 \mathrm{Ad_n})} = \phi^{-1} \frac{\exp \left(-0.25\alpha  {\rm Ad}_{\rm sh} \right)}{\exp (-0.12 \mathrm{Ad_n})}
\end{equation}
where $\phi$ is the volume fraction of particles. The numerator in Eq. (\ref{eq_fbr_ratio}) is rearranged to form an adhesion parameter ${\rm Ad}_{\rm sh}$, which measures the relative importance of adhesion and the shear stress. ${\rm Ad}_{\rm sh}$ has been successfully used to predict whether an agglomerate exposed to the simple shear flow will break or not \citep{RuanCES2020}. Given the parameters in our simulation conditions, we have $f_{br}^{sh}/f_{br}^{co} \ll 1$, indicating that shear-induced breakage can be neglected in the current work. However, increasing the effective shear rate ($G_{\rm{eff}}$) and decreasing the volume fraction ($\phi$) of the particles can both magnify the relative importance of shear-induced breakage. Given (\ref{eq_fbr_ratio}), it is straightforward to determine the dominant breakage mechanism. Our results extend those of \citet{SetoPRE2011,VanniLangmuir2011, FellayJCIS2012, BonaJFM2014,BablerJFM2015}, which focus on the breakage of agglomerates due to \textit{hydrodynamic stresses}, forming a more complete picture of breakage in turbulent flows.

In the present work, we have also shown that for adhesive particles with moderate inertia ($\rm{St}\approx 1$), the breakage events are more likely caused by particles ejected from strong vortices and happens in strain regions. It should be noted that the Reynolds number $\rm{Re}_{\lambda}$ currently used in the DNS-DEM simulation is fixed as $93$, which is a modest value. Higher $\rm{Re}_{\lambda}$ results in stronger intermittency and more intense vortex and strain structures, which give rise to extremely high impact velocities. Such intense structures, however, occupy smaller volumes as $\rm{Re}_{\lambda}$ increases \citep{PicardoPRF2019}. These competing effects would cause a nonmonotonic variation of the breakage rate. For heavy particles with $\rm{St} \ge 10$, the radial relative velocity increases with $\rm{Re}_{\lambda}$ since the particles carry a memory of more energetic motions as $\rm{Re}_{\lambda}$ increases. Such an effect is expected to increases the collision-induced breakage rate according to (\ref{eq_col_bdl2}). Other effects, including the correlated and extreme collision events \citep{bec2016abrupt, SawNC2016} and multifractal statistics of velocities differences \citep{saw2014extreme}, appear in high-Reynolds-number ﬂows may also contribute to the breakage rate. A complete picture of agglomeration and breakage, therefore, should include the role of both the turbulent transport and particle-level interactions, which will be systematically investigated in future studies.

\section*{Acknowledgements}
S.Q.L. acknowledges support from the National Natural Science Foundation of China (Grant No. 51725601). We are grateful to Prof. Federico Toschi at Technische Universiteit Eindhoven, Marshall, Prof. Jeff Marshall at Vermont and Prof. Eric Climent at Université de Toulouse for fruitful discussions, and Prof. Chao Sun and Mr Ruan at Tsinghua University for their useful suggestions.

\section*{Declaration of Interests}
The authors report no conflict of interest.

\bibliographystyle{jfm}
\bibliography{draft_breakage}

\begin{thebibliography}{78}
\expandafter\ifx\csname natexlab\endcsname\relax\def\natexlab#1{#1}\fi
\def\au#1{#1} \def\ed#1{#1} \def\yr#1{#1}\def\at#1{#1}\def\jt#1{\textit{#1}}
  \def\bt#1{#1}\def\bvol#1{\textbf{#1}} \def\vol#1{#1} \def\pg#1{#1}
  \def\publ#1{#1}\def\arxiv#1{#1}\def\org#1{#1}\def\st#1{\textit{#1}}

\bibitem[Ayala {\em et~al.\/}(2008)Ayala, Rosa \& Wang]{AyalaNJP2008}
{\sc \au{Ayala, Orlando}, \au{Rosa, Bogdan} \& \au{Wang, Lian-Ping}} \yr{2008}
  \at{Effects of turbulence on the geometric collision rate of sedimenting
  droplets. part 2. theory and parameterization}.  \jt{New Journal of Physics}
  \bvol{10}~(7),  \pg{075016}.

\bibitem[B{\"a}bler {\em et~al.\/}(2015)B{\"a}bler, Biferale, Brandt, Feudel,
  Guseva, Lanotte, Marchioli, Picano, Sardina, Soldati \&
  Toschi]{BablerJFM2015}
{\sc \au{B{\"a}bler, M.~U.}, \au{Biferale, L.}, \au{Brandt, L.}, \au{Feudel,
  U.}, \au{Guseva, K.}, \au{Lanotte, A.~S.}, \au{Marchioli, C.}, \au{Picano,
  F.}, \au{Sardina, G.}, \au{Soldati, A.} \& \au{Toschi, F.}} \yr{2015}
  \at{Numerical simulations of aggregate breakup in bounded and unbounded
  turbulent flows}.  \jt{Journal of Fluid Mechanics}  \bvol{766},
  \pg{104--128}.

\bibitem[B{\"a}bler \& Morbidelli(2008)]{bablerJFM2008}
{\sc \au{B{\"a}bler, M.~U.} \& \au{Morbidelli, M.and~BA{\L}DYGA, J.}} \yr{2008}
   \at{Modelling the breakup of solid aggregates in turbulent flows}.
  \jt{Journal of Fluid Mechanics}  \bvol{612},  \pg{261--289}.

\bibitem[Balachandar \& Eaton(2010)]{BalachandarAR2010}
{\sc \au{Balachandar, S} \& \au{Eaton, John~K}} \yr{2010}  \at{Turbulent
  dispersed multiphase flow}.  \jt{Annual review of fluid mechanics}
  \bvol{42},  \pg{111--133}.

\bibitem[Barnocky \& Davis(1988)]{BarnockyPOF1988}
{\sc \au{Barnocky, Guy} \& \au{Davis, Robert~H.}} \yr{1988}
  \at{Elastohydrodynamic collision and rebound of spheres: Experimental
  verification}.  \jt{Physics of Fluids}  \bvol{31}~(6),  \pg{1324}.

\bibitem[Bec {\em et~al.\/}(2013)Bec, Musacchio \& Ray]{BecPRE2013}
{\sc \au{Bec, J.}, \au{Musacchio, S.} \& \au{Ray, S.~S.}} \yr{2013}  \at{Sticky
  elastic collisions}.  \jt{Physical Review E}  \bvol{87}~(6),  \pg{063013}.

\bibitem[Bec {\em et~al.\/}(2016)Bec, Ray, Saw \& Homann]{bec2016abrupt}
{\sc \au{Bec, J.}, \au{Ray, S.~S.}, \au{Saw, E.~W.} \& \au{Homann, H.}}
  \yr{2016}  \at{Abrupt growth of large aggregates by correlated coalescences
  in turbulent flow}.  \jt{Physical Review E}  \bvol{93}~(3),  \pg{031102}.

\bibitem[Bhatnagar {\em et~al.\/}(2018)Bhatnagar, Gustavsson \&
  Mitra]{BhatnagarPRE2018}
{\sc \au{Bhatnagar, A.}, \au{Gustavsson, K.} \& \au{Mitra, D.}} \yr{2018}
  \at{Statistics of the relative velocity of particles in turbulent flows:
  Monodisperse particles}.  \jt{Physical Review E}  \bvol{97}~(2),
  \pg{023105}.

\bibitem[Chang {\em et~al.\/}(2017)Chang, Zheng, Yang, Fang, Gao, Luo \&
  Cen]{ChangFuel2017}
{\sc \au{Chang, Q.}, \au{Zheng, C.}, \au{Yang, Z.}, \au{Fang, M.}, \au{Gao,
  X.}, \au{Luo, Z.} \& \au{Cen, K.}} \yr{2017}  \at{Electric agglomeration
  modes of coal-fired fly-ash particles with water droplet humidification}.
  \jt{Fuel}  \bvol{200},  \pg{134--145}.

\bibitem[Chen {\em et~al.\/}(2019{\natexlab{{\em a\/}}})Chen, Li \&
  Marshall]{ChenPRF2019}
{\sc \au{Chen, S.}, \au{Li, S.~Q.} \& \au{Marshall, J.~S.}}
  \yr{2019{\natexlab{{\em a\/}}}}  \at{Exponential scaling in early-stage
  agglomeration of adhesive particles in turbulence}.  \jt{Physical Review
  Fluids}  \bvol{4}~(2),  \pg{024304}.

\bibitem[Chen {\em et~al.\/}(2015)Chen, Li \& Yang]{ChenPT2015}
{\sc \au{Chen, S.}, \au{Li, S.~Q.} \& \au{Yang, M.}} \yr{2015}
  \at{Sticking/rebound criterion for collisions of small adhesive particles:
  Effects of impact parameter and particle size}.  \jt{Powder Technology}
  \bvol{274},  \pg{431--440}.

\bibitem[Chen {\em et~al.\/}(2019{\natexlab{{\em b\/}}})Chen, Liu \&
  Li]{ChenCES2019}
{\sc \au{Chen, Sheng}, \au{Liu, Wenwei} \& \au{Li, Shuiqing}}
  \yr{2019{\natexlab{{\em b\/}}}}  \at{A fast adhesive discrete element method
  for random packings of fine particles}.  \jt{Chemical Engineering Science}
  \bvol{193},  \pg{336--345}.

\bibitem[Chen {\em et~al.\/}(2016)Chen, Liu \& Li]{ChenPRE2016}
{\sc \au{Chen, S.}, \au{Liu, W.} \& \au{Li, S.~Q.}} \yr{2016}  \at{Effect of
  long-range electrostatic repulsion on pore clogging during microfiltration}.
  \jt{Physical Review E}  \bvol{94}~(6),  \pg{063108}.

\bibitem[Chokshi {\em et~al.\/}(1993)Chokshi, Tielens \&
  Hollenbach]{ChokshiAJ1993}
{\sc \au{Chokshi, Arati}, \au{Tielens, AGGM} \& \au{Hollenbach, D}} \yr{1993}
  \at{Dust coagulation}.  \jt{The Astrophysical Journal}  \bvol{407},
  \pg{806--819}.

\bibitem[Davis {\em et~al.\/}(1986)Davis, Serayssol \& Hinch]{DavisJFM1986}
{\sc \au{Davis, Robert~H.}, \au{Serayssol, Jean-Marc} \& \au{Hinch, E.~J.}}
  \yr{1986}  \at{The elastohydrodynamic collision of two spheres}.  \jt{Journal
  of Fluid Mechanics}  \bvol{163},  \pg{479--497}.

\bibitem[De~Bona {\em et~al.\/}(2014)De~Bona, Lanotte \& Vanni]{BonaJFM2014}
{\sc \au{De~Bona, J.}, \au{Lanotte, A.~S.} \& \au{Vanni, M.}} \yr{2014}
  \at{Internal stresses and breakup of rigid isostatic aggregates in
  homogeneous and isotropic turbulence}.  \jt{Journal of Fluid Mechanics}
  \bvol{755},  \pg{365--396}.

\bibitem[Di~Felice(1994)]{di1994}
{\sc \au{Di~Felice, R.}} \yr{1994}  \at{The voidage function for fluid-particle
  interaction systems}.  \jt{International Journal of Multiphase Flow}
  \bvol{20}~(1),  \pg{153--159}.

\bibitem[Dizaji {\em et~al.\/}(2019)Dizaji, Marshall \& Grant]{Dizaji2019JFM}
{\sc \au{Dizaji, F.~F.}, \au{Marshall, J.~S.} \& \au{Grant, J.~R.}} \yr{2019}
  \at{Collision and breakup of fractal particle agglomerates in a shear flow}.
  \jt{Journal of Fluid Mechanics}  \bvol{862},  \pg{592--623}.

\bibitem[Dong {\em et~al.\/}(2018)Dong, Mei, Li, Shang \& Li]{DongPT2018}
{\sc \au{Dong, M.}, \au{Mei, Y.}, \au{Li, X.}, \au{Shang, Y.} \& \au{Li, S.}}
  \yr{2018}  \at{Experimental measurement of the normal coefficient of
  restitution of micro-particles impacting on plate surface in different
  humidity}.  \jt{Powder technology}  \bvol{335},  \pg{250--257}.

\bibitem[Ernst {\em et~al.\/}(2013)Ernst, Dietzel \& Sommerfeld]{ErnstAM2013}
{\sc \au{Ernst, M.}, \au{Dietzel, M.} \& \au{Sommerfeld, M.}} \yr{2013}  \at{A
  lattice boltzmann method for simulating transport and agglomeration of
  resolved particles}.  \jt{Acta Mechanica}  \bvol{224}~(10),  \pg{2425--2449}.

\bibitem[Falkovich {\em et~al.\/}(2002)Falkovich, Fouxon \&
  Stepanov]{FalkovichNature2002}
{\sc \au{Falkovich, G}, \au{Fouxon, A} \& \au{Stepanov, MG}} \yr{2002}
  \at{Acceleration of rain initiation by cloud turbulence}.  \jt{Nature}
  \bvol{419}~(6903),  \pg{151}.

\bibitem[Fang {\em et~al.\/}(2019)Fang, Wang, Zhang, Wei, Wu \&
  Sun]{FangJAS2019}
{\sc \au{Fang, Z.}, \au{Wang, H.}, \au{Zhang, Y.}, \au{Wei, M.}, \au{Wu, X.} \&
  \au{Sun, L.}} \yr{2019}  \at{A finite element method ({F}{E}{M}) study on
  adhesive particle-wall normal collision}.  \jt{Journal of Aerosol Science}
  \bvol{134},  \pg{80--94}.

\bibitem[Fellay \& Vanni(2012)]{FellayJCIS2012}
{\sc \au{Fellay, L.~S.} \& \au{Vanni, M.}} \yr{2012}  \at{The effect of flow
  configuration on hydrodynamic stresses and dispersion of low density rigid
  aggregates}.  \jt{Journal of Colloid and Interface Science}  \bvol{388}~(1),
  \pg{47--55}.

\bibitem[Flesch {\em et~al.\/}(1999)Flesch, Spicer \&
  Pratsinis]{flesch1999laminar}
{\sc \au{Flesch, J.~C.}, \au{Spicer, P.~T.} \& \au{Pratsinis, S.~E.}} \yr{1999}
   \at{Laminar and turbulent shear-induced flocculation of fractal aggregates}.
   \jt{AIChE Journal}  \bvol{45}~(5),  \pg{1114--1124}.

\bibitem[Gu {\em et~al.\/}(2016)Gu, Ozel \& Sundaresan]{gu2016modified}
{\sc \au{Gu, Yile}, \au{Ozel, Ali} \& \au{Sundaresan, Sankaran}} \yr{2016}
  \at{A modified cohesion model for cfd--dem simulations of fluidization}.
  \jt{Powder Technology}  \bvol{296},  \pg{17--28}.

\bibitem[Higashitani {\em et~al.\/}(2001)Higashitani, Iimura \&
  Sanda]{HigashitaniCES2001}
{\sc \au{Higashitani, K.}, \au{Iimura, K.} \& \au{Sanda, H.}} \yr{2001}
  \at{Simulation of deformation and breakup of large aggregates in flows of
  viscous fluids}.  \jt{Chemical Engineering Science}  \bvol{56}~(9),
  \pg{2927--2938}.

\bibitem[Iimura {\em et~al.\/}(2009)Iimura, Suzuki, Hirota \&
  Higashitani]{IimuraAPT2009}
{\sc \au{Iimura, K.}, \au{Suzuki, M.}, \au{Hirota, M.} \& \au{Higashitani, K.}}
  \yr{2009}  \at{Simulation of dispersion of agglomerates in gas
  phase--acceleration field and impact on cylindrical obstacle}.  \jt{Advanced
  Powder Technology}  \bvol{20}~(2),  \pg{210--215}.

\bibitem[Israelachvili(2011)]{Israelachvili2011}
{\sc \au{Israelachvili, J~N}} \yr{2011} {\em Intermolecular and surface
  forces\/}.  \publ{Academic press}.

\bibitem[Jaworek {\em et~al.\/}(2018)Jaworek, Marchewicz, Sobczyk, Krupa \&
  Czech]{JaworekPECS2018}
{\sc \au{Jaworek, A.}, \au{Marchewicz, A.}, \au{Sobczyk, A.~T.}, \au{Krupa, A.}
  \& \au{Czech, T.}} \yr{2018}  \at{Two-stage electrostatic precipitators for
  the reduction of pm2. 5 particle emission}.  \jt{Progress in Energy and
  Combustion Science}  \bvol{67},  \pg{206--233}.

\bibitem[Jiang \& Logan(1991)]{JiangEST1991}
{\sc \au{Jiang, Q.} \& \au{Logan, B.~E.}} \yr{1991}  \at{Fractal dimensions of
  aggregates determined from steady-state size distributions}.
  \jt{Environmental Science \& Technology}  \bvol{25}~(12),  \pg{2031--2038}.

\bibitem[Johnson {\em et~al.\/}(1971)Johnson, Kendall \& Roberts]{JKR}
{\sc \au{Johnson, KL}, \au{Kendall, K} \& \au{Roberts, AD}} \yr{1971}
  \at{Surface energy and the contact of elastic solids}.  \jt{Proc. R. Soc.
  Lond. A}  \bvol{324}~(1558),  \pg{301--313}.

\bibitem[Jones(2005)]{Jones2005}
{\sc \au{Jones, T.~B.}} \yr{2005} {\em Electromechanics of particles\/}.
  \publ{Cambridge University Press}.

\bibitem[Kellogg {\em et~al.\/}(2017)Kellogg, Liu, LaMarche \&
  Hrenya]{KelloggJFM2017}
{\sc \au{Kellogg, K.~M.}, \au{Liu, P.}, \au{LaMarche, C.~Q.} \& \au{Hrenya,
  C.~M.}} \yr{2017}  \at{Continuum theory for rapid cohesive-particle flows:
  general balance equations and discrete-element-method-based closure of
  cohesion-specific quantities}.  \jt{Journal of Fluid Mechanics}  \bvol{832},
  \pg{345--382}.

\bibitem[Krijt {\em et~al.\/}(2013)Krijt, G{\"u}ttler, Hei{\ss}elmann, Dominik
  \& Tielens]{krijt2013energy}
{\sc \au{Krijt, S.}, \au{G{\"u}ttler, C.}, \au{Hei{\ss}elmann, D.},
  \au{Dominik, C.} \& \au{Tielens, A. G. G.~M.}} \yr{2013}  \at{Energy
  dissipation in head-on collisions of spheres}.  \jt{Journal of Physics D:
  Applied Physics}  \bvol{46}~(43),  \pg{435303}.

\bibitem[Li \& Marshall(2007)]{LiJAS2007}
{\sc \au{Li, S.~Q.} \& \au{Marshall, J.~S.}} \yr{2007}  \at{Discrete element
  simulation of micro-particle deposition on a cylindrical fiber in an array}.
  \jt{Journal of Aerosol Science}  \bvol{38}~(10),  \pg{1031--1046}.

\bibitem[Li {\em et~al.\/}(2011)Li, Marshall, Liu \& Yao]{LiPECS2011}
{\sc \au{Li, S.~Q.}, \au{Marshall, J.~S.}, \au{Liu, G.} \& \au{Yao, Q.}}
  \yr{2011}  \at{Adhesive particulate flow: The discrete-element method and its
  application in energy and environmental engineering}.  \jt{Progress in Energy
  and Combustion Science}  \bvol{37}~(6),  \pg{633--668}.

\bibitem[Liu \& Hrenya(2018)]{LiuPRL2018}
{\sc \au{Liu, P.} \& \au{Hrenya, C.~M.}} \yr{2018}  \at{Cluster-induced
  deagglomeration in dilute gravity-driven gas-solid flows of cohesive grains}.
   \jt{Physical Review Letters}  \bvol{121}~(23),  \pg{238001}.

\bibitem[Liu {\em et~al.\/}(2017)Liu, Jin, Chen, Makse \& Li]{LiuSM2017}
{\sc \au{Liu, W.}, \au{Jin, Y.}, \au{Chen, S.}, \au{Makse, H.~A.} \& \au{Li,
  S.~Q.}} \yr{2017}  \at{Equation of state for random sphere packings with
  arbitrary adhesion and friction}.  \jt{Soft Matter}  \bvol{13}~(2),
  \pg{421--427}.

\bibitem[Liu {\em et~al.\/}(2015)Liu, Li, Baule \& Makse]{LiuSM2015}
{\sc \au{Liu, W.}, \au{Li, S.~Q.}, \au{Baule, A.} \& \au{Makse, H.~A.}}
  \yr{2015}  \at{Adhesive loose packings of small dry particles}.  \jt{Soft
  Matter}  \bvol{11}~(32),  \pg{6492--6498}.

\bibitem[Liu \& Wu(2019)]{LiuPOF2019}
{\sc \au{Liu, W.} \& \au{Wu, C.~Y.}} \yr{2019}  \at{Analysis of inertial
  migration of neutrally buoyant particle suspensions in a planar poiseuille
  flow with a coupled lattice boltzmann method-discrete element method}.
  \jt{Physics of Fluids}  \bvol{31}~(6),  \pg{063301}.

\bibitem[Liu \& Wu(2020)]{LiuAICHE2020}
{\sc \au{Liu, Wenwei} \& \au{Wu, Chuan-Yu}} \yr{2020}  \at{Migration and
  agglomeration of adhesive micro-particle suspensions in a pressure-driven
  duct flow}.  \jt{AIChE Journal}  \pg{p. e16974}.

\bibitem[Lu {\em et~al.\/}(2010)Lu, Nordsiek, Saw \& Shaw]{lu2010clustering}
{\sc \au{Lu, J.}, \au{Nordsiek, H.}, \au{Saw, E.~W.} \& \au{Shaw, R.~A.}}
  \yr{2010}  \at{Clustering of charged inertial particles in turbulence}.
  \jt{Physical Review Letters}  \bvol{104}~(18),  \pg{184505}.

\bibitem[Lu \& Shaw(2015)]{lu2015charged}
{\sc \au{Lu, J.} \& \au{Shaw, R.~A.}} \yr{2015}  \at{Charged particle dynamics
  in turbulence: Theory and direct numerical simulations}.  \jt{Physics of
  Fluids}  \bvol{27}~(6),  \pg{065111}.

\bibitem[Marshall(2009)]{MarshallJCP2009}
{\sc \au{Marshall, J.~S.}} \yr{2009}  \at{Discrete-element modeling of
  particulate aerosol flows}.  \jt{Journal of Computational Physics}
  \bvol{228}~(5),  \pg{1541--1561}.

\bibitem[Marshall(2011)]{Marshall2011}
{\sc \au{Marshall, J.~S.}} \yr{2011}  \at{Viscous damping force during head-on
  collision of two spherical particles}.  \jt{Physics of Fluids}
  \bvol{23}~(1),  \pg{013305}.

\bibitem[Marshall \& Li(2014)]{Marshall2014}
{\sc \au{Marshall, J.~S.} \& \au{Li, S.~Q.}} \yr{2014} {\em Adhesive Particle
  Flow\/}.  \publ{Cambridge University Press}.

\bibitem[Pan \& Padoan(2010)]{PanJFM2010}
{\sc \au{Pan, L.} \& \au{Padoan, P.}} \yr{2010}  \at{Relative velocity of
  inertial particles in turbulent flows}.  \jt{Journal of Fluid Mechanics}
  \bvol{661},  \pg{73--107}.

\bibitem[Peng {\em et~al.\/}(2019)Peng, Ayala \& Wang]{PengJFM2019}
{\sc \au{Peng, C.}, \au{Ayala, O.~M.} \& \au{Wang, L.~P.}} \yr{2019}  \at{A
  direct numerical investigation of two-way interactions in a particle-laden
  turbulent channel flow}.  \jt{Journal of Fluid Mechanics}  \bvol{875},
  \pg{1096--1144}.

\bibitem[Picardo {\em et~al.\/}(2019)Picardo, Agasthya, Govindarajan \&
  Ray]{PicardoPRF2019}
{\sc \au{Picardo, J.~R.}, \au{Agasthya, L.}, \au{Govindarajan, R.} \& \au{Ray,
  S.~S.}} \yr{2019}  \at{Flow structures govern particle collisions in
  turbulence}.  \jt{Physical Review Fluids}  \bvol{4}~(3),  \pg{032601}.

\bibitem[Pumir \& Wilkinson(2016)]{pumir2016}
{\sc \au{Pumir, A.} \& \au{Wilkinson, M.}} \yr{2016}  \at{Collisional
  aggregation due to turbulence}.  \jt{Annual Review of Condensed Matter
  Physics}  \bvol{7},  \pg{141--170}.

\bibitem[Renault {\em et~al.\/}(2009)Renault, Sancey, Charles, Morin-Crini,
  Badot, Winterton \& Crini]{RenaultCEJ2009}
{\sc \au{Renault, F.}, \au{Sancey, B.}, \au{Charles, J.}, \au{Morin-Crini, N.},
  \au{Badot, P.~M.}, \au{Winterton, P.} \& \au{Crini, G.}} \yr{2009}
  \at{Chitosan flocculation of cardboard-mill secondary biological wastewater}.
   \jt{Chemical Engineering Journal}  \bvol{155}~(3),  \pg{775--783}.

\bibitem[Royer {\em et~al.\/}(2009)Royer, Evans, Oyarte, Guo, Kapit,
  M{\"o}bius, Waitukaitis \& Jaeger]{RoyerNature2009}
{\sc \au{Royer, J.~R.}, \au{Evans, D.~J.}, \au{Oyarte, L.}, \au{Guo, Q.},
  \au{Kapit, E.}, \au{M{\"o}bius, M.~E.}, \au{Waitukaitis, S.~R.} \&
  \au{Jaeger, H.~M.}} \yr{2009}  \at{High-speed tracking of rupture and
  clustering in freely falling granular streams}.  \jt{Nature}
  \bvol{459}~(7250),  \pg{1110}.

\bibitem[Ruan {\em et~al.\/}(2020)Ruan, Chen \& Li]{RuanCES2020}
{\sc \au{Ruan, X.}, \au{Chen, S.} \& \au{Li, S.~Q.}} \yr{2020}  \at{Structural
  evolution and breakage of dense agglomerates in shear flow and taylor-green
  vortex}.  \jt{Chemical Engineering Science}  \bvol{211},  \pg{115261}.

\bibitem[Rubinow \& Keller(1961)]{RubinowJFM1961}
{\sc \au{Rubinow, S.~I.} \& \au{Keller, J.~B.}} \yr{1961}  \at{The transverse
  force on a spinning sphere moving in a viscous fluid}.  \jt{Journal of Fluid
  Mechanics}  \bvol{11}~(3),  \pg{447--459}.

\bibitem[Saffman(1965)]{SaffmanJFM1965}
{\sc \au{Saffman, P.~G.}} \yr{1965}  \at{The lift on a small sphere in a slow
  shear flow}.  \jt{Journal of Fluid Mechanics}  \bvol{22}~(2),  \pg{385--400}.

\bibitem[Saffman \& Turner(1956)]{SaffmanJFM1956}
{\sc \au{Saffman, P.~G.} \& \au{Turner, J.~S.}} \yr{1956}  \at{On the collision
  of drops in turbulent clouds}.  \jt{Journal of Fluid Mechanics}
  \bvol{1}~(1),  \pg{16--30}.

\bibitem[Salazar \& Collins(2012)]{SalazarJFM2012}
{\sc \au{Salazar, J.~P.L.C.} \& \au{Collins, L.~R.}} \yr{2012}  \at{Inertial
  particle relative velocity statistics in homogeneous isotropic turbulence}.
  \jt{Journal of Fluid Mechanics}  \bvol{696},  \pg{45--66}.

\bibitem[Saw {\em et~al.\/}(2014)Saw, Bewley, Bodenschatz, Ray \&
  Bec]{saw2014extreme}
{\sc \au{Saw, E.~W.}, \au{Bewley, G.~P.}, \au{Bodenschatz, E.}, \au{Ray, S.~S.}
  \& \au{Bec, J.}} \yr{2014}  \at{Extreme fluctuations of the relative
  velocities between droplets in turbulent airflow}.  \jt{Physics of Fluids}
  \bvol{26}~(11),  \pg{111702}.

\bibitem[Saw {\em et~al.\/}(2016)Saw, Kuzzay, Faranda, Guittonneau, Daviaud,
  Wiertel-Gasquet, Padilla \& Dubrulle]{SawNC2016}
{\sc \au{Saw, E.~W.}, \au{Kuzzay, D.}, \au{Faranda, D.}, \au{Guittonneau, A.},
  \au{Daviaud, F.}, \au{Wiertel-Gasquet, C.}, \au{Padilla, V.} \& \au{Dubrulle,
  B.}} \yr{2016}  \at{Experimental characterization of extreme events of
  inertial dissipation in a turbulent swirling flow}.  \jt{Nature
  communications}  \bvol{7},  \pg{12466}.

\bibitem[Saw {\em et~al.\/}(2008)Saw, Shaw, Ayyalasomayajula, Chuang \&
  Gylfason]{SawPRL2008}
{\sc \au{Saw, E.~W.}, \au{Shaw, R.~A.}, \au{Ayyalasomayajula, S.}, \au{Chuang,
  P.~Y.} \& \au{Gylfason, A.}} \yr{2008}  \at{Inertial clustering of particles
  in high-reynolds-number turbulence}.  \jt{Physical Review Letters}
  \bvol{100}~(21),  \pg{214501}.

\bibitem[Seto {\em et~al.\/}(2011)Seto, Botet \& Briesen]{SetoPRE2011}
{\sc \au{Seto, R.}, \au{Botet, R.} \& \au{Briesen, H.}} \yr{2011}
  \at{Hydrodynamic stress on small colloidal aggregates in shear flow using
  stokesian dynamics}.  \jt{Physical Review E}  \bvol{84}~(4),  \pg{041405}.

\bibitem[Squires \& Eaton(1991)]{SquiresPOF1991}
{\sc \au{Squires, K.~D.} \& \au{Eaton, J.~K.}} \yr{1991}  \at{Preferential
  concentration of particles by turbulence}.  \jt{Physics of Fluids A: Fluid
  Dynamics}  \bvol{3}~(5),  \pg{1169--1178}.

\bibitem[Steinpilz \& Wurm(2019)]{SteinpilzNP2019}
{\sc \au{Steinpilz, T., Joeris K. Jungmann F. Wolf D. Brendel L. Teiser J.
  Shinbrot~T.} \& \au{Wurm, G.}} \yr{2019}  \at{Electrical charging overcomes
  the bouncing barrier in planet formation}.  \jt{Nature Physics} .

\bibitem[S{\"u}mer \& Sitti(2008)]{SumerJAST2008}
{\sc \au{S{\"u}mer, B.} \& \au{Sitti, M.}} \yr{2008}  \at{Rolling and spinning
  friction characterization of fine particles using lateral force microscopy
  based contact pushing}.  \jt{Journal of Adhesion Science and Technology}
  \bvol{22}~(5-6),  \pg{481--506}.

\bibitem[Sundaram \& Collins(1997)]{SundaramJFM1997}
{\sc \au{Sundaram, S.} \& \au{Collins, L.~R.}} \yr{1997}  \at{Collision
  statistics in an isotropic particle-laden turbulent suspension. part 1.
  direct numerical simulations}.  \jt{Journal of Fluid Mechanics}  \bvol{335},
  \pg{75--109}.

\bibitem[Tagawa {\em et~al.\/}(2012)Tagawa, Mercado, Prakash, Calzavarini, Sun
  \& Lohse]{TagawaJFM2012}
{\sc \au{Tagawa, Y.}, \au{Mercado, J.~M.}, \au{Prakash, V.~N.},
  \au{Calzavarini, E.}, \au{Sun, C.} \& \au{Lohse, D.}} \yr{2012}
  \at{Three-dimensional lagrangian vorono{\"\i} analysis for clustering of
  particles and bubbles in turbulence}.  \jt{Journal of fluid mechanics}
  \bvol{693},  \pg{201--215}.

\bibitem[Tsuji {\em et~al.\/}(1992)Tsuji, Tanaka \& Ishida]{TsujiPT1992}
{\sc \au{Tsuji, Yutaka}, \au{Tanaka, Toshitsugu} \& \au{Ishida, T}} \yr{1992}
  \at{Lagrangian numerical simulation of plug flow of cohesionless particles in
  a horizontal pipe}.  \jt{Powder Technology}  \bvol{71}~(3),  \pg{239--250}.

\bibitem[Vanni \& Gastaldi(2011)]{VanniLangmuir2011}
{\sc \au{Vanni, M.} \& \au{Gastaldi, A.}} \yr{2011}  \at{Hydrodynamic forces
  and critical stresses in low-density aggregates under shear flow}.
  \jt{Langmuir}  \bvol{27}~(21),  \pg{12822--12833}.

\bibitem[Voss \& Finlay(2002)]{VossIJP2002}
{\sc \au{Voss, A.} \& \au{Finlay, W.~H.}} \yr{2002}  \at{Deagglomeration of dry
  powder pharmaceutical aerosols}.  \jt{International Journal of Pharmaceutics}
   \bvol{248}~(1-2),  \pg{39--50}.

\bibitem[Vo{\ss}kuhle {\em et~al.\/}(2013)Vo{\ss}kuhle, L{\'e}v{\^e}que,
  Wilkinson \& Pumir]{VosskuhlePRE2013}
{\sc \au{Vo{\ss}kuhle, M.}, \au{L{\'e}v{\^e}que, E.}, \au{Wilkinson, M.} \&
  \au{Pumir, A.}} \yr{2013}  \at{Multiple collisions in turbulent flows}.
  \jt{Physical Review E}  \bvol{88}~(6),  \pg{063008}.

\bibitem[Wang {\em et~al.\/}(2019)Wang, Wan, Liu \& Wang]{WangCES2019}
{\sc \au{Wang, G.}, \au{Wan, D.and~Peng, C.}, \au{Liu, K.} \& \au{Wang, L.~P.}}
  \yr{2019}  \at{Lbm study of aggregation of monosized spherical particles in
  homogeneous isotropic turbulence}.  \jt{Chemical Engineering Science}
  \bvol{201},  \pg{201--211}.

\bibitem[Wang {\em et~al.\/}(2000)Wang, Wexler \& Zhou]{WangJFM2000}
{\sc \au{Wang, L.~P.}, \au{Wexler, A.~S.} \& \au{Zhou, Y.}} \yr{2000}
  \at{Statistical mechanical description and modelling of turbulent collision
  of inertial particles}.  \jt{Journal of Fluid Mechanics}  \bvol{415},
  \pg{117--153}.

\bibitem[Wei {\em et~al.\/}(2019)Wei, Zhang, Fang, Wu \& Sun]{WeiNED2019}
{\sc \au{Wei, M.}, \au{Zhang, Y.}, \au{Fang, Z.}, \au{Wu, X.} \& \au{Sun, L.}}
  \yr{2019}  \at{Graphite aerosol release to the containment in a water ingress
  accident of high temperature gas-cooled reactor (htgr)}.  \jt{Nuclear
  Engineering and Design}  \bvol{342},  \pg{170--175}.

\bibitem[Wilkinson {\em et~al.\/}(2006)Wilkinson, Mehlig \&
  Bezuglyy]{WilkinsonPRL2006}
{\sc \au{Wilkinson, M.}, \au{Mehlig, B.} \& \au{Bezuglyy, V.}} \yr{2006}
  \at{Caustic activation of rain showers}.  \jt{Physical Review Letters}
  \bvol{97}~(4),  \pg{048501}.

\bibitem[Xiong {\em et~al.\/}(2019)Xiong, Li, Fei \& Luo]{XiongIJMF2019}
{\sc \au{Xiong, Y.}, \au{Li, J.}, \au{Fei, F.and~Liu, Z.} \& \au{Luo, W.}}
  \yr{2019}  \at{Influence of coherent vortex structures in subgrid scale
  motions on particle statistics in homogeneous isotropic turbulence}.
  \jt{International Journal of Multiphase Flow}  \bvol{113},  \pg{358--370}.

\bibitem[Yang \& Hunt(2006)]{YangPOF2006}
{\sc \au{Yang, F.~L.} \& \au{Hunt, M.~L.}} \yr{2006}  \at{Dynamics of
  particle-particle collisions in a viscous liquid}.  \jt{Physics of Fluids}
  \bvol{18}~(12),  \pg{121506}.

\bibitem[Yang {\em et~al.\/}(2013)Yang, Li \& Yao]{YangPT2013}
{\sc \au{Yang, M.}, \au{Li, S.~Q.} \& \au{Yao, Q.}} \yr{2013}  \at{Mechanistic
  studies of initial deposition of fine adhesive particles on a fiber using
  discrete-element methods}.  \jt{Powder Technology}  \bvol{248},  \pg{44--53}.

\bibitem[Zhou {\em et~al.\/}(2001)Zhou, Wexler \& Wang]{ZhouJFM2001}
{\sc \au{Zhou, Y.}, \au{Wexler, A.~S.} \& \au{Wang, L.~P.}} \yr{2001}
  \at{Modelling turbulent collision of bidisperse inertial particles}.
  \jt{Journal of Fluid Mechanics}  \bvol{433},  \pg{77--104}.

\end{thebibliography}

\end{document}